\documentclass[10pt,twocolumn]{IEEEtran}
\usepackage[T1]{fontenc}
\ifCLASSINFOpdf
\else
\fi

\usepackage{dblfloatfix}
\usepackage{cite}
\usepackage{wrapfig}
\usepackage{graphicx}
\usepackage{url}

\usepackage{multicol}
\usepackage{amsmath}
\usepackage{amsbsy}
\usepackage{amssymb}
\usepackage{amsthm}
\usepackage{graphics}
\usepackage{epsfig}
\usepackage{flushend}
\usepackage{graphicx}
\usepackage{psfrag}
\usepackage{color}
\usepackage[config, labelfont=small, textfont=small]{caption,subfig}
\usepackage{fancyhdr}
\usepackage{dsfont}
\usepackage{bm}
\newtheorem{proposition}{Proposition}
\newtheorem{theorem}{Theorem}
\newtheorem{lemma}{Lemma}

\newtheorem{remark}{Remark}
\newtheorem{definition}{Definition}
\newtheorem{corollary}{Corollary}
\newtheorem{conjecture}{Conjecture}

\allowdisplaybreaks
\enlargethispage{\baselineskip}

\let\OLDthebibliography\thebibliography
\renewcommand\thebibliography[1]{
  \OLDthebibliography{#1}
  \setlength{\parskip}{0pt}
  \setlength{\itemsep}{0pt plus 0.3ex}
}

\usepackage{dblfloatfix}
\usepackage{balance}

\usepackage{url}
\definecolor{title}{RGB}{180,0,0}
\definecolor{other}{RGB}{171,0,255}
\definecolor{name}{RGB}{255,0,0}
\definecolor{phd}{RGB}{0,0,240}

\def\R{\mathbb{R}}

\def\smallint{\begingroup\textstyle \int\endgroup}

\begin{document}
\title{{{Cox Models for Vehicular Networks: \\ SIR Performance and Equivalence}}}
\author{\IEEEauthorblockN{Jeya Pradha Jeyaraj, \textit{Student Member, IEEE,} and Martin Haenggi, \textit{Fellow, IEEE}}  \\
\thanks{The authors are with the Department of Electrical Engineering, University of Notre Dame, Notre Dame, IN 46556, USA. (e-mail: $\lbrace$jjeyaraj, mhaenggi$\rbrace$@nd.edu). 

This work was supported in part by Toyota Motor North America R\&D and by the U.S.~National Science Foundation (grant CCF 2007498).
}
}

\maketitle 
\begin{abstract}
We introduce a general framework for the modeling and analysis of vehicular networks by defining street systems as random 1D subsets of $\mathbb{R}^{2}$. The street system, in turn, specifies the random intensity measure of a Cox process of vehicles, {\em{i.e.,}} vehicles form independent 1D Poisson point processes on each street. Models in this Coxian framework can characterize streets of different lengths and orientations forming intersections or T-junctions. The lengths of the streets can be infinite or finite and mutually independent or dependent. We analyze the reliability of communication for different models, where reliability is the probability that a vehicle at an intersection,  a T-junction, or a general location can receive a message successfully from a transmitter at a certain distance. Further, we introduce a notion of equivalence between vehicular models, which means that a representative model can be used as a proxy for other models in terms of reliability. Specifically, we prove that the Poisson stick process-based vehicular network is equivalent to the Poisson line process-based and Poisson lilypond model-based vehicular networks, and their rotational variants.

\end{abstract}


\section{Introduction}
\label{sg}
\subsection{Motivation and Related Work}
{ Through vehicle-to-vehicle (V2V) communication, vehicles can broadcast the information required for safety such as their speed, brake status, position, etc., to other vehicles. This can alert the vehicles about the events happening in their vicinity that even the best sensors installed in the vehicles may fail to anticipate. Further, infrastructure nodes such as roadside units, smart traffic lights, etc., can inform the drivers of the road/traffic conditions through vehicle-to-infrastructure (V2I) communication. High reliability is a key requirement for safety-based V2V and V2I applications due to their mission-critical nature. Reliability or success probability is defined as the probability that a broadcast message is received reliably at a certain distance. The factors that affect the reliability include the locations of the transmitting and receiving vehicles and interfering vehicles, and wireless medium. }

{ An analysis of GPS traces of taxis in Beijing and Porto in~\cite{cui} demonstrates that (i) the taxi locations cannot be modeled as random points as in a 2D Poisson point process (PPP) neglecting the street geometry as claimed in~\cite{jiang}, and (ii) it is desirable to model the intersections, which are critical for vehicle safety. Limiting the reliability analysis to a single or a small number of streets is certainly useful in its own right, albeit insufficient to obtain insights into the network behavior at intersections and T-junctions~\cite{jeong,  blaz, koufos3} and in dense urban scenarios~\cite{erik}. Hence, street geometry is integral to vehicular network modeling.
 
The street systems in different geographical regions may differ in their structure, street lengths, and degree of regularity. The authors in~\cite{centrality} have studied the street systems of 18 cities in different parts of the world. They divide the street systems into two classes---(i) self-organized patterns that are formed historically without the control of any central agency, and (ii) planned regular grid-like patterns. Cities such as Ahmedabad (India), Cairo (Egypt), and Venice (Italy) are examples of self-organized patterns with unimodal street length distributions. Los Angeles, Richmond, and San Francisco in the United States are examples of grid-like patterns with multimodal street length distributions. It is worth noting that that even in the cities containing grid-like street patterns, the street lengths are finite and vary significantly.

In the literature~\cite{dhillon, morlot, choi2, chetlur_recent, chetlur, choi, sial}, street systems are most commonly modeled as a random collection of lines with uniform orientations using Poisson line processes (PLPs). While assuming that all streets are infinitely long may lead to useful results, it either overestimates the total interference or underestimates the local density of vehicles. For example, a street in a city that is 2 km long may have a density of 50 cars per km. But extending it to an infinite street would mean such a high car density extends to very remote rural regions far outside the city, which is not realistic. Further, in the PLP, each pair of streets forms an intersection and there are no T-junctions. Fig.~\ref{fig:rome} shows a part of Rome as an example, where few streets are long enough to be approximated by infinite streets, and there exist many T-junctions.
\begin{figure}
\centering
{\includegraphics[scale=0.15]{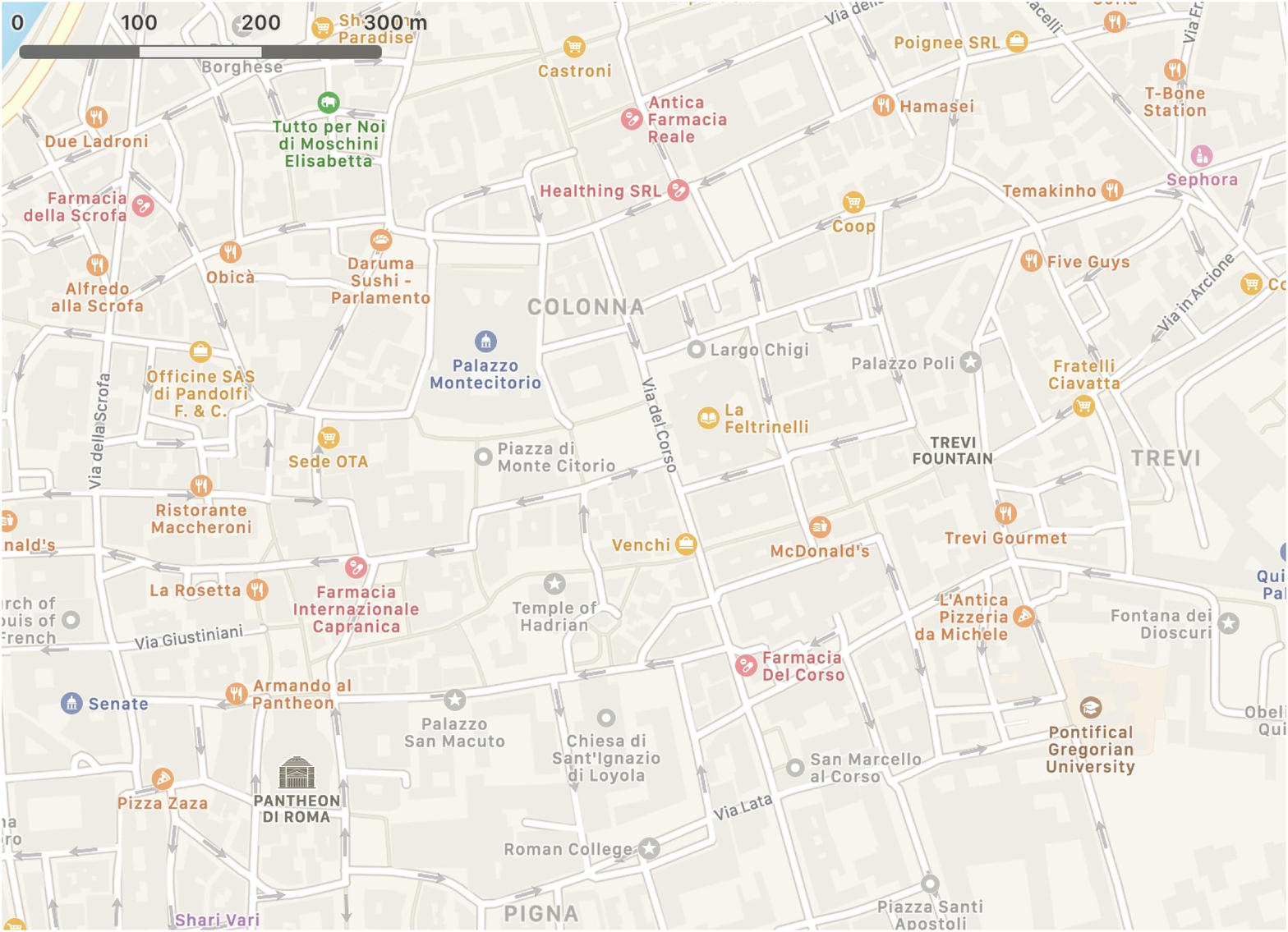}} 
\caption{Part of Rome's city center. Visual inspection shows a mean street length of about 250 m and many T-junctions.}
\label{fig:rome}
\end{figure}

Consequently, we need street models that can characterize finite variations in the street lengths, and intersections and T-junctions. The edges of Poisson-Voronoi tessellation, Poisson-Delaunay tessellation, and Poisson line tessellation are considered to model streets of finite lengths in~\cite{voss}. Estimators for the probability densities of the inter-node distances are derived for these tessellations owing to their intractability. A simpler alternative is to use line segments or sticks as we show later. In this work, we introduce a general framework for the modeling and analysis of vehicular networks. Under this framework, we develop models that represent street systems varying from regular grid-like patterns to irregular hodgepodges and characterize the uncertainty in the vehicle locations on the streets. In particular, the street lengths can be infinitely long or varying finitely and mutually independent or dependent resulting in intersections or T-junctions.}

An important question is whether we need a different spatial model for each region. Alternatively, does there exist an equivalence between the models such that a single model is sufficient to analyze two different regions? If yes, a representative subset can be used to analyze a larger set of models, reducing the computation time and costs associated with network design and planning. We show that some models developed within the framework are equivalent, in a precise sense defined later. We use the mathematical toolsets from stochastic geometry~\cite{haenggi} to model and analyze random spatial patterns. 

\subsection{Contributions}
We present a general framework for the modeling and analysis of vehicular networks, where the lengths of the streets can be infinite or finite, mutually independent or dependent, and the street orientations can have different levels of regularity. Lines or line segments (sticks) characterize the streets. The street geometries of the spatial models that can be characterized in this framework include but not limited to 1) the orthogonal grid with exponential spacing (OG) whose lines are of infinite length with orthogonal orientations, 2) the PLP whose lines are of infinite length with irregular orientations, 3) the Poisson stick process (PSP) whose sticks can have random lengths and orientations, and 4) the Poisson lilypond model (PLM) whose sticks grow in a random direction until they touch another stick, mimicking the growth of lilies in a pond.

The OG, PLP, and PSP inherently form intersections, whereas in the PLM, the dependence between the sticks results in T-junctions. To the best of our knowledge, this is the first work to model and analyze vehicular networks with street systems formed by line segments involving intersections or T-junctions.

The vehicle locations on each street independently form 1D PPPs, {\em{i.e.,}} the spatial models in the presented framework are Cox vehicular networks. Our contributions are:
\begin{itemize}
\item We introduce a street system partitioning approach that decomposes a street system into points of different orders, where the order quantifies the number of street segments covering the point. This decomposition permits a unified approach to the performance analysis of vehicles at general locations, intersections, and T-junctions.
\item We derive the nearest-neighbor distance distribution in the PSP-based vehicular network, and we show that the nearest-neighbor distance distribution in the PLM-based vehicular network can be tightly approximated by that in the PSP-based vehicular network.
\item We derive analytical expressions for the probability that a vehicle at a general location, an intersection, or a T-junction successfully receives from its transmitter located at a certain distance, in each of the considered models.
\item We introduce a notion of equivalence between the models based on the success probability. The expression for the success probability of the typical vehicle in the PSP-based vehicular network can be used to evaluate that in the vehicular networks formed by other models by mapping the parameters such as street and vehicle intensities, and street length distribution. We show that the equivalence also holds under random link distances.
\item We show that the Cox vehicular networks behave like a 2D PPP in the low-reliability regime. In contrast, in the high-reliability regime, the success probability of the typical vehicle in the Cox vehicular networks tends to that in the network formed only by the street(s) that the typical vehicle belongs to. This corroborates earlier findings~\cite{cui} that the vehicles cannot be generally modeled as 2D PPPs and that the street geometry plays an important role. 

\end{itemize}
 
\section{System Model} 
Throughout the paper, we will use the definitions and notations presented in this section unless otherwise stated. Let $b(x,r)$ denote a disk of radius $r$ centered at $x$, and $o \triangleq (0,0)$ denote the origin. Let $\vert \cdot \vert_{d}$ denote the Lebesgue measure in $d$ dimensions.

\subsection{General Framework}
\begin{definition}[Street System]
\label{street_system}
{{A street system $\mathcal{S}$ is a stationary random closed subset of $\mathbb{R}^{2}$ with $\vert \mathcal{S} \vert_2 = 0$ that contains no singletons or isolated points. Due to the stationarity, 
 \begin{equation}
\mathbb{E}\vert \mathcal{S} \cap \mathcal{B} \vert_1 = \tau \vert \mathcal{B} \vert_2 \hspace{3mm} \text{for Borel sets} \hspace{1mm} \mathcal{B} \subset \mathbb{R}^2, 
\label{eq:street}
\end{equation}
where $\tau$ is the mean total street length per unit area.}}
\end{definition}
$\vert \mathcal{S} \vert_2 = 0$ in Definition~\ref{street_system} implies that $\mathcal{S}$ is a random 1D subset of the plane consisting of lines, line segments or sticks, curved segments or arcs, that characterize the streets. Let $\Xi=\{\xi_1,\xi_2,\ldots\}$ be a collection of 1D subsets in $\R^2$ such that for $i\neq j$, $|\xi_i\cap \xi_j|_1=0$ and $\xi_i\cup \xi_j$ is not a 1D subset. The street system $\mathcal{S}$ is the union of 1D subsets, {\em{i.e.,}}
\begin{equation}
\mathcal{S} \triangleq \bigcup_{\xi \in\Xi} \xi.
\end{equation}
 The elements of $\mathcal{S}$ are points in $\mathbb{R}^2$ but those of $\Xi$ are 1D subsets, which are sets themselves, and hence $\mathcal{S} \neq \Xi$. Further, $\mathcal{S}$ uniquely characterizes $\Xi$ and vice-versa, {\em{i.e.,}} there exists a one-to-one correspondence between $\mathcal{S}$ and $\Xi$.
Any street system $\mathcal{S}$ can be partitioned as follows.
\begin{definition}[Street System Decomposition]
Let $\mathcal{P}_m \triangleq \lbrace{z \in \mathbb{R}^{2} : \vert \mathcal{S}\cap b(z,r)\vert_1 \sim mr, r \to 0 \rbrace}$ denote the set of points of order $m \in \mathbb{N}$ in the street system $\mathcal{S}$. 
\end{definition}
The sets $\mathcal{P}_m$ are disjoint, and their union equals $\mathcal{S}$, {\em{i.e.,}} $ \lbrace \mathcal{P}_m \rbrace_{m \in \mathbb{N}}$ is a partition of $\mathcal{S}$. For $m\neq 2$, the sets $\mathcal{P}_m$ are countable and form simple and stationary point processes. $\mathcal{P}_2$ is the only set with $\vert\mathcal{P}_2\vert_1 > 0$, in fact, $|\mathcal{P}_2|_1=\infty$. We have $\mathcal{P}_2=\mathcal{S}$ almost everywhere, {\em{i.e.,}} $\mathcal{P}_2$ is an open set and $\mathcal{S}=\mathrm{cl}(\mathcal{P}_2)$, where $\mathrm{cl}$ denotes the closure.
$\mathcal{P}_1$ are endpoints, $\mathcal{P}_3$ are T-junctions, $\mathcal{P}_4$ are intersections, $\mathcal{P}_5$ are intersections with a T-junction, $\mathcal{P}_6$ are three-way intersections (three streets intersecting at one point), etc. 
Let 
\begin{equation}
\mathcal{M} \triangleq \lbrace m \in \mathbb{N}: \mathbb{P}(\mathcal{P}_m = \emptyset) = 0 \rbrace
\label{eq:index_set}
\end{equation}
denote the index set of the non-empty components. Then $\mathcal{S} = \bigcup_{m \in \mathcal{M}} \mathcal{P}_m$ is called an $\mathcal{M}-$indexed street system.
Using $\mathcal{M}$, we can categorize different street systems. For example, a $(2,4)-$street system refers to a street geometry without endpoints and T-junctions but with intersections.

\begin{definition}[Vehicular Point Process]
A vehicular point process $\mathcal{V} \subset \mathbb{R}^2$ is a Cox process with random intensity measure $\Upsilon(\mathcal{B})=\lambda |\mathcal{S}\cap \mathcal{B}|_1$.
\end{definition}
Equivalently, $\Upsilon(\mathcal{B})=\lambda |\mathcal{P}_2\cap \mathcal{B}|_1$ since a.s. $\mathcal{V} \subset \mathcal{P}_2$. This implies that the vehicles form independent 1D PPPs on each street. By Definition~1, the 2D intensity measure of $\mathcal{V}$ is $\mathbb{E}[\Upsilon(\mathcal{B})]=\lambda \mathbb{E}[|\mathcal{S}\cap \mathcal{B}|_1] = \lambda \tau\vert\mathcal{B}\vert_2$.

\subsection{Vehicular Network Models}
Here, we present a few models that fall under our framework. 
 A street system may include curves or circles. In this work, we focus only on street systems formed by lines or sticks. 

A line (an infinitely long street) $L$ can be represented as 
\begin{equation}
L({x},{\varphi}) = \lbrace{(a,b) \in \mathbb{R}^{2} : a\cos\varphi + b\sin\varphi = x \rbrace},
\label{eq:line}
\end{equation}
where $(\vert x \vert, \varphi)$ are the polar coordinates of the foot of the perpendicular from the origin $o$ to $L$.  

\begin{definition}[Line-based Poisson Street System] Consider a marked point process on $\mathbb{R} \times \Omega$, whose ground process is a PPP $\Phi_1$ of intensity $\mu$ on $\mathbb{R}$ and marks are i.i.d. on $\Omega  \subseteq [0, \pi)$. Let $\nu$ denote the distribution of $\Omega$. The collection of lines $\Xi_{\mathrm{L}} = \lbrace{ L(x, \varphi): (x, \varphi) \in \Phi_1 \times \Omega\rbrace}$ with the intensity measure $ \Lambda_{\Xi}(\mathrm{d}x\hspace{0.3mm} \mathrm{d}\varphi) = \mu \hspace{0.3mm}\mathrm{d}x \times \mathrm{d} \nu(\varphi)$ 
forms
\begin{itemize}
\item an \textit{orthogonal grid with exponential spacing} (OG) if $ \mathrm{d} \nu(\varphi) = 0.5 \delta_{0} + 0.5 \delta_{\pi/2}$,
\item a \textit{Poisson line process} (PLP) if $ \mathrm{d} \nu(\varphi) = \mathrm{d}\varphi/\pi$.
\end{itemize} 
The line-based Poisson street system is $\mathcal{S} = \bigcup_{L \in \Xi_{\mathrm{L}}} L $.
\label{def:line_system}
\end{definition}

The OG and PLP inherently form intersections and hence are classified as $(2,4)-$street systems. Streets of varying finite lengths can be represented using sticks. A stick $S$ is defined by its midpoint $y$, orientation $\varphi$, and half-length $h$, and represented as a closed set as
\begin{align}
S(y, \varphi, h) = [y-h u(\varphi), y+h u(\varphi)], 
\label{eq:segment}
\end{align}
where $u(\varphi) = (\cos \varphi, \sin\varphi)$. Alternatively, $S(y, \varphi, h) = y + {\mathrm{rot}}_{\varphi}([-h,h])$, where $\mathrm{rot}_{\varphi}$ denotes the rotation by $\varphi$ around $o$. Now, we are ready to define the  stick-based street system.

\begin{definition}[Stick-based Poisson Street System]
\label{def:stick_system}
Let $\mathcal{Q} = \lbrace (y_i, \boldsymbol{t}_i) \rbrace$, $i \in \mathbb{N}$,  denote an i.i.d. marked point process with the ground process $\lbrace y_i \rbrace$ forming a 2D PPP $\Phi_2$ of intensity $\mu$. Associate with $y_i$ a 2D mark $\boldsymbol{t}_i = (\varphi_i, h_i)$, where the orientation $\varphi_i$ is i.i.d. on $[0, \pi)$, and $h_i$ is the half-length. 
The countable collection of sticks $\Xi_{\mathrm{S}} = \lbrace S(y_i, \varphi_i, h_i) \rbrace $
forms 
\begin{itemize}
\item a \textit{Poisson stick process} (PSP) if the half-lengths are i.i.d. with some distribution $F_{H}$.
\item a \textit{Poisson lilypond model} (PLM){\footnote{The PLM is denoted as LM-I in~\cite{Daley}. We note that the other model, LM-II in~\cite{Daley}, has similar properties as the PLM.}} if each stick grows from zero length at a constant rate on both sides until one of its endpoints hits another stick, thereby forming a T-junction. 
\end{itemize}
Then $\mathcal{S} = \bigcup_{S \in \Xi_{\mathrm{S}}} S$ defines the stick-based Poisson street system.
\end{definition}
Let $y_i \equiv(\gamma_i,\phi_i)$ in polar coordinates, where $\gamma_i \in \mathbb{R}^{+}$, and $\phi_i \in [0, 2\pi)$. The PSP has no T-junctions a.s., thus forming a $(1,2,4)-$street system, whereas the PLM has no intersections a.s. due to its touch-and-stop growth mechanism, thus forming a $(1,2,3)-$street system.

We only consider the street systems containing points of order up to 4. The analyses shown in this work can be extended to higher-order street systems as well.

\begin{figure*}
\centering
\subfloat[$\mu = 1$ and $\lambda = 0.1$]{%
\includegraphics[scale = 0.56]{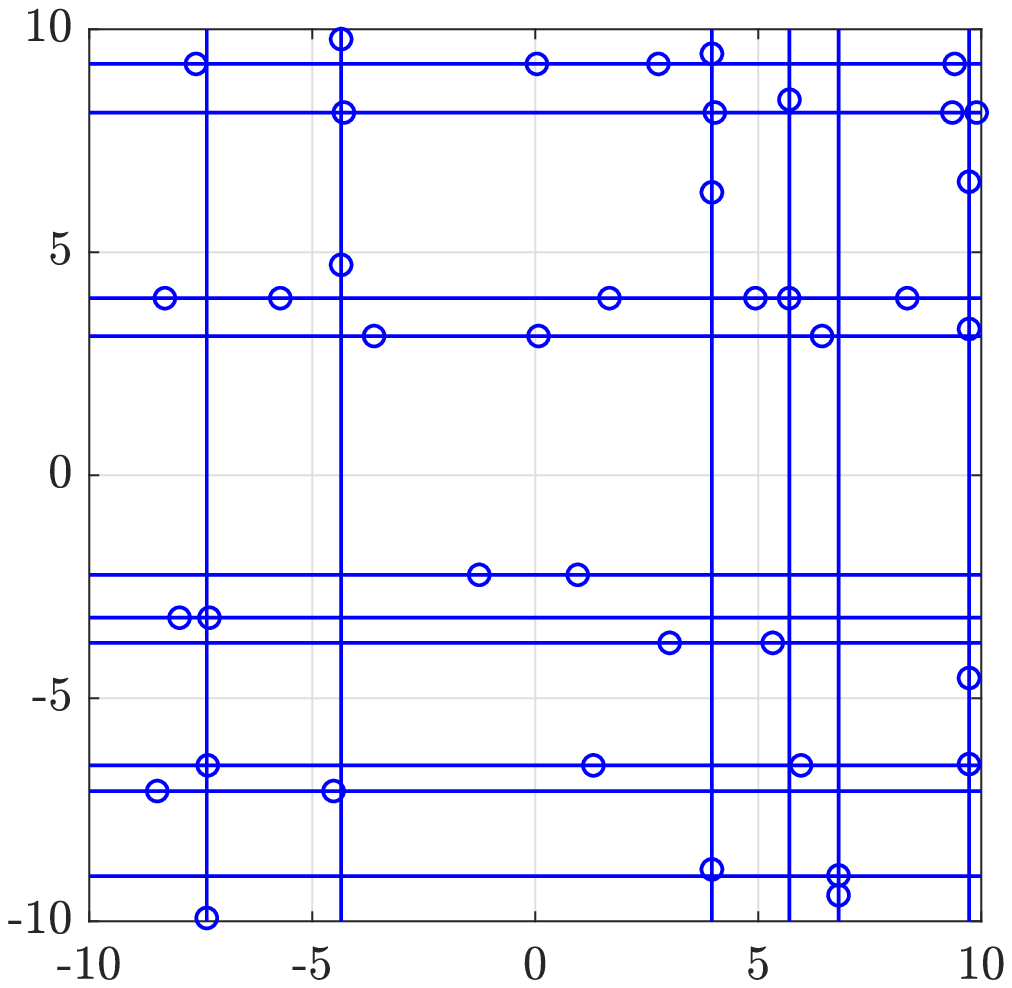}
\label{fig:og_ppp}
 }\hspace{25mm}
 \vspace{3mm}
\subfloat[$\mu = 1$ and $\lambda = 0.1$]{%
\includegraphics[scale = 0.56] {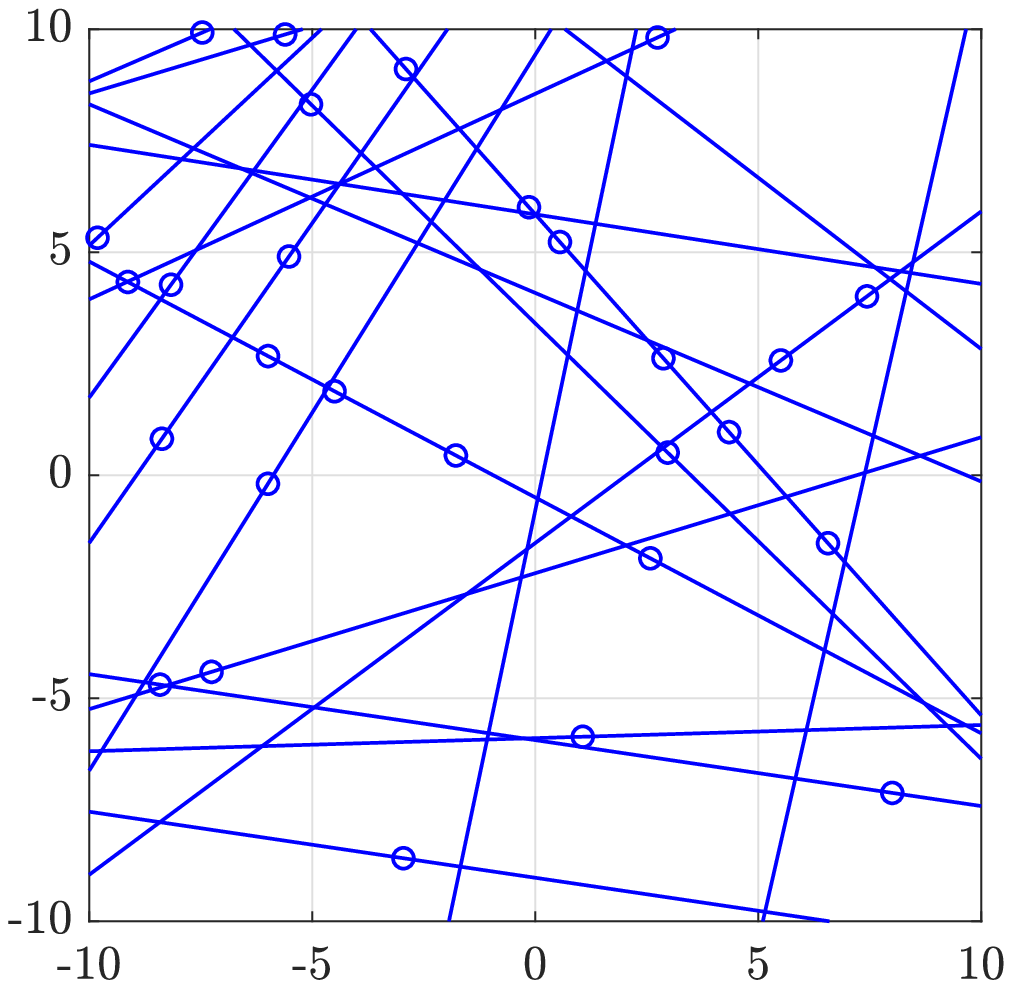}
\label{fig:plp_ppp}
 }\hspace{25mm}
\subfloat[$\mu = 0.1$ and $\lambda = 0.1$]{%
\includegraphics[scale = 0.56]{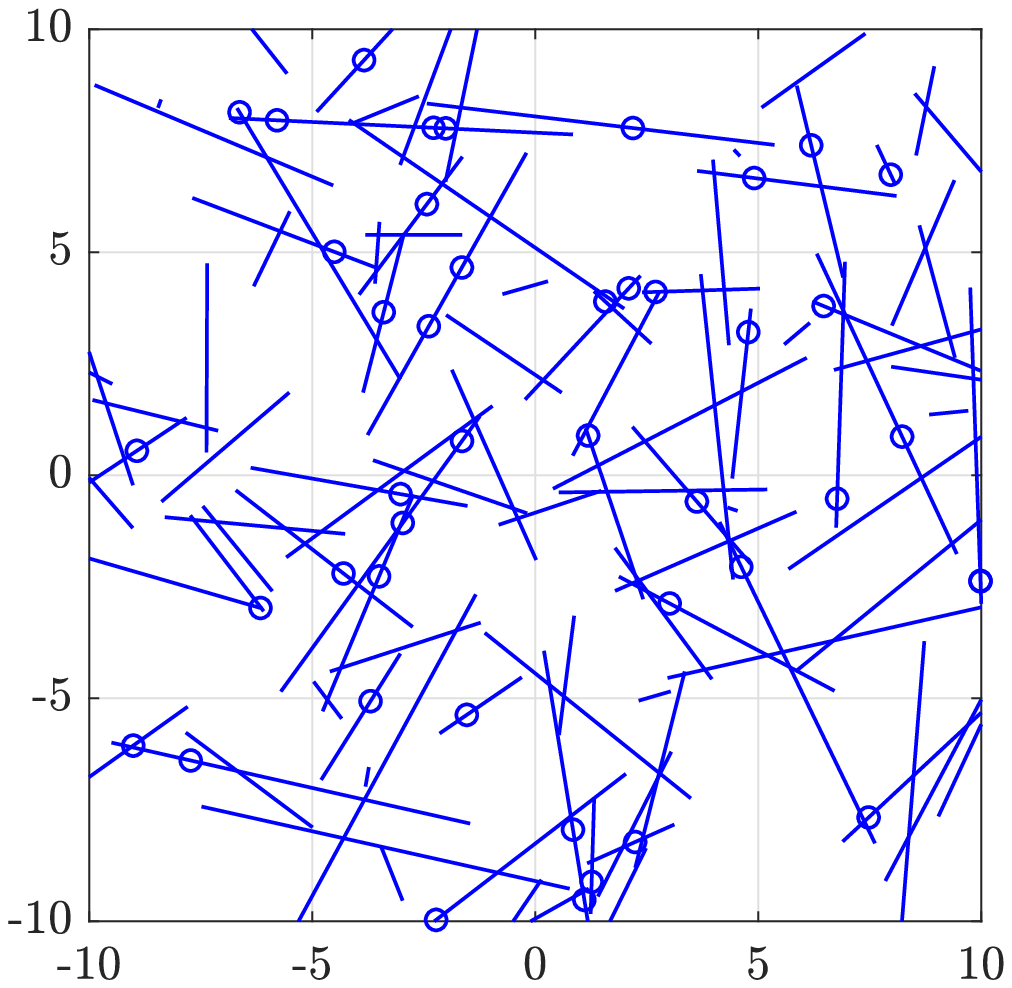}
\label{fig:lsp_ppp}
}\hspace{26mm}
\subfloat[$\mu = 0.1$ and $\lambda = 0.1$]{%
\includegraphics[scale = 0.56]{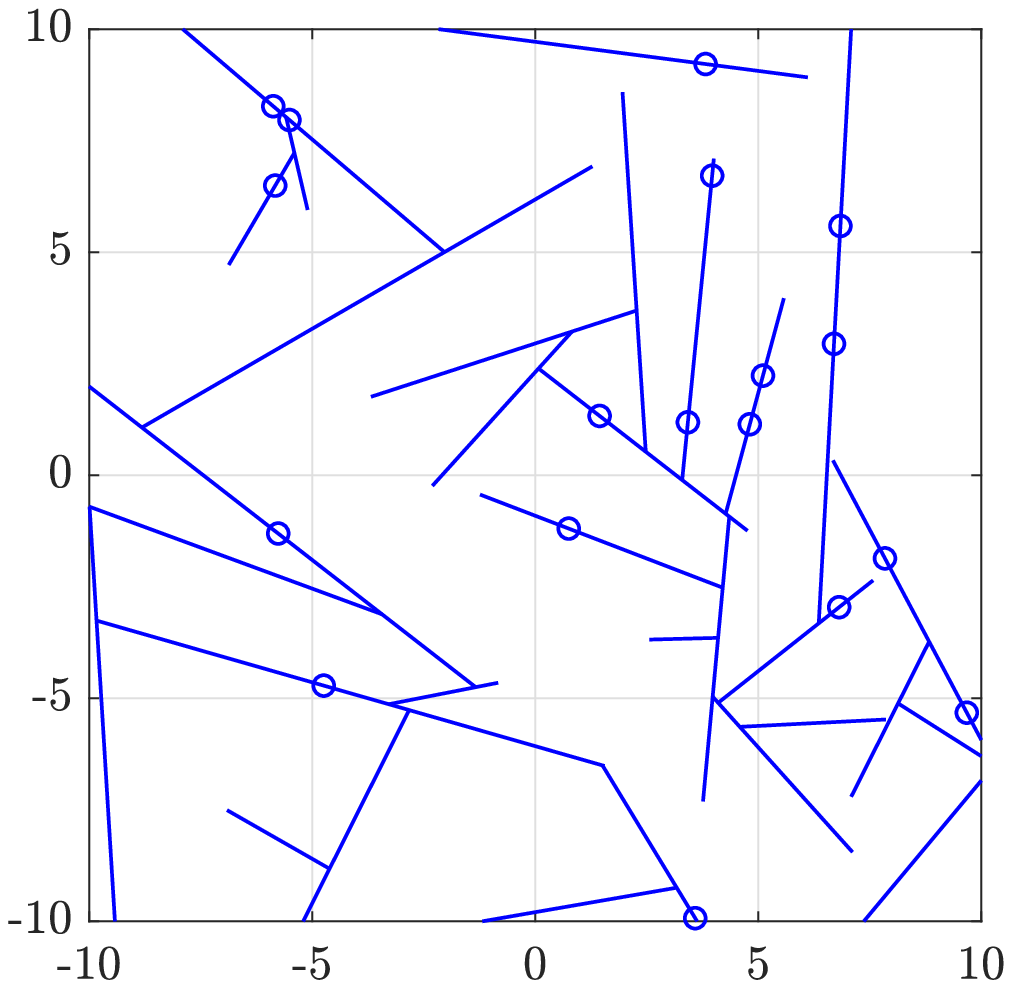}
\label{fig:lp_ppp}
}
\caption{\label{fig:models} Snapshots of vehicular networks: (a) OG-PPP (b) PLP-PPP (c) PSP-PPP and (d) PLM-PPP. Lines or sticks denote the streets, and `$\circ$' denote the vehicles. {{The stick lengths in (c) are Rayleigh distributed with parameter 2.}} }
\end{figure*}

We refer to the vehicular point processes formed by the OG, PLP, PSP, and PLM as the OG-PPP, PLP-PPP, PSP-PPP, and PLM-PPP, respectively. Fig.~\ref{fig:models} depicts sample realizations.

\subsection{Performance Metric and Types of Vehicles}
Each vehicle broadcasts with probability $p$ following the slotted ALOHA protocol. Then the intensity of active transmitters on each street in each time slot is $\lambda p$. Our metric of interest is the success probability or reliability, which is the probability of a vehicle successfully receiving the message from a transmitter at distance $D$. If a transmitter can communicate to a receiver at a distance $D$, then the other receivers within distance $D$ are also highly likely to receive the message. The transmitter can be another vehicle, a roadside unit, a pedestrian, or any other node.

To define a meaningful network-wide metric, we focus on the success probability of a representative vehicle (receiver) whose performance corresponds to the average of that of all vehicles. In point process theory, this representative vehicle is called `the typical point.' In our context, it is `the typical vehicle.' As vehicles are located on the street(s), having a vehicle at the origin implies that at least one street passes through the origin. Under expectation over the vehicular point process, a vehicle conditioned to be at the origin becomes the typical vehicle. Note that we can condition the typical vehicle to be at any location since the vehicular network is stationary owing to the underlying stationary street system (Definition 1) and the homogeneity of the PPP. The typical vehicle's transmitter is assumed to be at a distance $D$ from the origin.

The performance of vehicular communication at intersections and T-junctions is crucial as they are more prone to accidents. In view of this, we evaluate the success probabilities of three kinds of vehicles: (i) \textit{the typical general vehicle} whose order is 2; (ii) \textit{the typical intersection vehicle} whose order is 4; and (iii) \textit{the typical T-junction vehicle} whose order is 3. The term \textit{typical vehicle} refers to all the three types of typical vehicles unless otherwise stated. Mathematically, the success probability of the typical vehicle of order $m$ at $o$ is defined as
\begin{equation}
    p_{m}(\theta)= \mathbb{P}({\sf{SIR}} > \theta \mid o \in \mathcal{P}_{m}), \hspace{2mm}  m = 2, 3, 4, 
    \label{eq:sir_def}
\end{equation}
where SIR is the signal-to-interference ratio measured at $o$ and $\theta$ parametrizes the target rate. The SIR for the typical vehicle with its transmitter at distance $D$ is given by
\begin{equation}
{\sf SIR} = \frac{g D^{-\alpha}}{\sum_{z \in \mathcal{V}}  g_{z}\Vert z \Vert^{-\alpha} \mathit{B}_z},
\label{eq:sir_expr}
\end{equation}
where $I = \sum_{z \in \mathcal{V}} g_{z}\Vert z \Vert^{-\alpha}$ is the total interference power at the origin. The channel power gains $g$ and $g_{z}$ are exponentially distributed with mean 1 (Rayleigh fading), and $\alpha $ is the path-loss exponent. $(\mathit{B}_z)_{z \in \mathcal{V}}$ is an i.i.d. sequence of Bernoulli random variables with mean $p$, the probability that the transmitter $z $ is active. 
Equipped with the performance metric, we next define the equivalence of spatial models.
\subsection{Equivalence}
\begin{definition}[Equivalence]
\label{def_equiv}
Two spatial models $\mathrm{A}$ and $\mathrm{B}$ are said to be $\epsilon$-equivalent with respect to the typical vehicle of order $m$ if the total variation distance of their SIR distributions is at most $\epsilon$, i.e., $\max \limits_{\theta} \vert p_{m}^{\mathrm{A}}(\theta) - p_{m}^{\mathrm{B}}(\theta) \vert = \epsilon$, $0 \leq \epsilon \leq 1$. 
If this holds with $\epsilon=0$, we call them strictly equivalent. Model $\mathrm{A}$ is said to be asymptotically equivalent to model $\mathrm{B}$ in the lower regime of $\theta$ if $(1-p_{m}^{\mathrm{A}}(\theta))/(1-p_{m}^{\mathrm{B}}(\theta)) \to 1$ as $\theta \to 0$, and in the higher regime of $\theta$ if $p_{m}^{\mathrm{A}}(\theta)/p_{m}^{\mathrm{B}}(\theta) \to 1$ as $\theta \to \infty$.{\footnote{{For all models, the success probability of the typical vehicle $p_{m}(\theta) \to 1$ as $\theta \to 0$. Hence, considering ratios of success probabilities is not meaningful in this regime as they would all be equivalent. Similarly, the outage probability $1-p_{m}(\theta) \to 1$ as $\theta\to\infty$, so in this regime, it does not make sense to consider the ratio of outage probabilities. To obtain non-trivial equivalence results, the quantities of interest need to go to zero.}}}
\end{definition}
If the street system $\mathrm{A}$ has index set $\mathcal{M}_{\mathrm{A}}$ and the street system $\mathrm{B}$ has index set $\mathcal{M}_{\mathrm{B}}$, then the equivalence of $\mathrm{A}$ and $\mathrm{B}$ is defined only for the typical vehicles of orders $\mathcal{M}_{\mathrm{A}} \cap \mathcal{M}_{\mathrm{B}}$. If $\mathrm{A}$ and $\mathrm{B}$ are strictly equivalent, then either $\mathrm{A}$ or $\mathrm{B}$ is sufficient to capture all the geographical regions that can be characterized by them. Else, substituting one model for the other would depend on the complexity of the model and the value of $\epsilon$, which we will discuss in detail in Section~\ref{section: plm_psp_equi}. 
 
\subsection{Further Notation}
\label{sec:notation}
Conditioning on $o \in \mathcal{P}_2$ implies that a street passes through the origin, conditioning on $o \in \mathcal{P}_3$ implies that a street passes through the origin while another street ends at the origin, and conditioning on $o \in \mathcal{P}_4$ implies that two streets pass through the origin. We denote by $\mathcal{V}^{m} = (\mathcal{V} \mid o \in \mathcal{P}_{m}) $ the point process of vehicles with the origin being on at least one street. Also, let $\mathcal{V}_{o}^{m}$ denote the vehicles on the streets that pass
through or end at the origin. $\mathcal{V}_{!}$ denotes the vehicles on the rest of the streets, {\em{i.e.,}} $\mathcal{V}_{!} = \mathcal{V}^{m} \setminus \mathcal{V}_{o}^{m}$. { We index the streets based on the distances of the perpendiculars from the streets to the origin.} Based on the indexing, $V_{k}$ denotes the point process of vehicles on the $k$th street. Then $\displaystyle \mathcal{V}_{o}^{m} = \cup_{1 \leq k \leq \lceil m/2 \rceil} V_{k}$, and $\displaystyle \mathcal{V}_{!} = \cup_{k > \lceil m/2 \rceil} V_{k}$.
Let $I_{o}^{m}$ and $I_{!}$ denote the interference from the vehicles in $\mathcal{V}_o^{m}$ and $\mathcal{V}_{!}$, respectively. $\delta \triangleq 2/\alpha$, where $\alpha$ is the path-loss exponent.

\section{Properties of Vehicular Networks}
\begin{lemma}
The mean total street length per unit area in the OG and PLP is $\tau = \mu$. 
\label{lemma_2d_og}
\end{lemma}

\begin{IEEEproof}
The total expected length of the lines that intersect $b(o,r)$ is given by
\begin{align}
\mathbb{E}[\vert \mathcal{S} \cap b(o,r) \vert_1] &= \mu \int_{0}^{\pi}\int_{-r}^{r} 2\sqrt{r^2-u^2} \hspace{0.5mm} \mathrm{d}u \hspace{0.5mm}  \mathrm{d} v (\varphi)  \nonumber \\
& = \mu \hspace{0.3mm} \vert b(o,r)\vert_2 . \label{eq:lemma1_IEEEproof}
\end{align}
By Definition 1 and \eqref{eq:lemma1_IEEEproof}, we get $\tau = \mu$. 
\end{IEEEproof}

\begin{lemma}[{{{\cite{parker}, Eq. 9}}}]
\label{lemma_lsp_2d}
The mean total stick length per unit area in the PSP is $\tau = 2 \mu \mathbb{E}[H]$.
\end{lemma}

Let $f_{H}(h)$ denote the probability density function (PDF) of the half-lengths of the sticks.
\begin{lemma}
\label{lemma_ins_paradox}
The half-lengths of the sticks that pass through the typical vehicle are distributed with density $\tilde{f}_{H}(h) = h f_{H}(h)/\mathbb{E}[H]$. 
\end{lemma}

Lemma~\ref{lemma_ins_paradox} is a case of the inspection paradox~\cite{bias}. The length of the stick that passes through the typical vehicle is biased by the fact that the mean number of points on the stick is proportional to its length. As a result, the half-lengths of those sticks follow the density function $\tilde{f}_{H}(h)$. { For example, consider a case where half the streets are of length $10^{-2}$ and the rest are of length $10^{2}$, and the intensity of vehicles on each street is 1. Then the typical vehicle lies on a long street with a probability of about $99.99\%$, which is very different from the $50\%$ probability of the typical street to be a long one. }
On the other hand, in the PLM, a stick that ends at a T-junction follows the inherent distribution $f_{H}(h)$ since each stick has exactly one such endpoint a.s. If all the sticks are of the same length $2h_0$, then $\tilde{f}_{H}(h) = f_{H}(h) = \delta(h-h_0)$. 

In the lilypond model, the growth of a stick is determined by the locations and the orientations of the other sticks since each stick grows at a constant rate until one of its endpoints hits another stick. This simultaneous touch-and-stop growth process makes it difficult, most likely impossible, to derive the exact distribution of the half-lengths. However, a suitable approximation for the distribution of the half-lengths can be found, as stated in Lemma~\ref{hl_approx}.

\begin{lemma}
\label{hl_approx}
The half-lengths of the sticks in the PLM are approximately Rayleigh distributed with PDF $\hat{f}_{H}(h) = 2 b h \exp(-bh^{2})$, $h \geq 0$. It follows that ${\mathbb{E}}[H] \approx \sqrt{\pi/4b}$, where $b$ is proportional to the street intensity $\mu$ and can be estimated from the empirical mean of the half-lengths.
\end{lemma}
\begin{IEEEproof}
See Appendix~\ref{appendix:hl_approx_proof}.
\end{IEEEproof}

\begin{figure}
\centering
\subfloat[]{\includegraphics[scale=0.57]{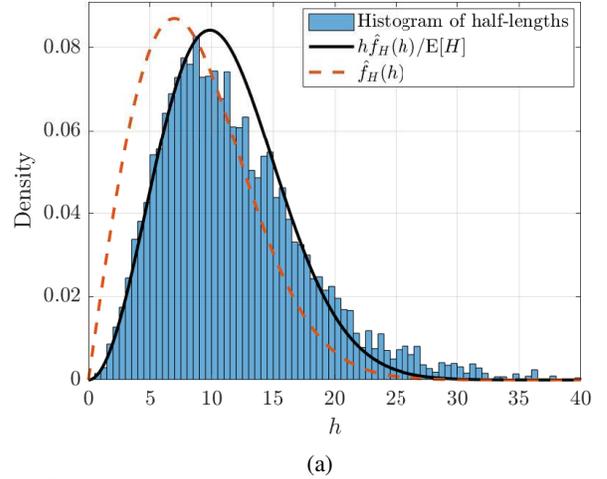} \label{fig:fitted_lm}}\hspace{10mm}
\subfloat[]{\includegraphics[scale=0.572]{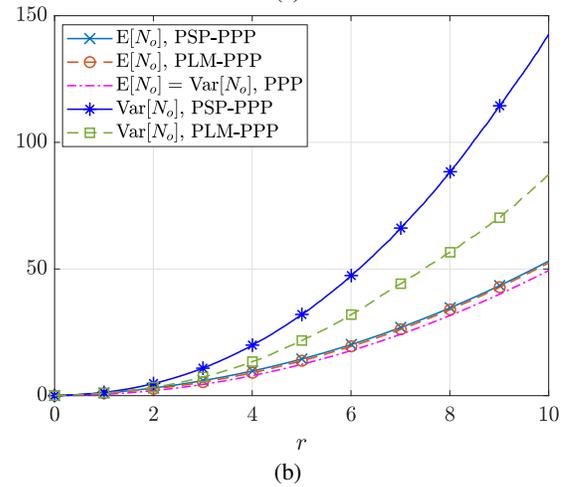}\label{fig:lsp_lm_var_mean}}
\caption{(a) Fitting $h \hat{f}_{H}(h)/\mathbb{E}[H]$ to the half-lengths of the streets that pass through the typical vehicle in the PLM. (b) Mean and variance of number of neighbors to the typical general vehicle in the PLM-PPP vs. PSP-PPP with $f_{H}(h) = \hat{f}_{H}(h)= 2 b h \exp(-bh^{2})$. $\mu = 0.01$ and $\lambda = 0.3$. The value of $b$ corresponding to $\mu = 0.01$ is 0.0103. The intensity of 2D PPP is $2 \lambda \mu \mathbb{E}[H]$, which is the 2D intensity of the PLM-PPP/PSP-PPP.}
\end{figure}

\setcounter{equation}{9}
\begin{figure*}[b]
\hrule
\begin{align}
F^{\mathrm{PSP-PPP}}_{R}(r)  = & 1 - \bigg[\int \limits_{0}^{\infty} \frac{1}{h}\int \limits_{0}^{h} \exp(-\lambda \ell(\gamma, 0, 0) )   \tilde{f}_{H}(h)  \hspace{0.3mm}  \mathrm{d}\gamma  \hspace{0.3mm}  \mathrm{d}h \bigg]^{ m/2 } \nonumber \\
& \times \exp\bigg(\hspace{-1mm}-\frac{\mu}{\pi}\int \limits_{0}^{\infty} \int \limits_{0}^{\pi} \int \limits_{0}^{2\pi} \int \limits_{0}^{r+h}   \exp(-\lambda \ell(\gamma, \phi, \varphi)) \gamma  {f}_{H}(h) \hspace{0.3mm}\mathrm{d}\gamma \hspace{0.3mm}  \mathrm{d}\phi \hspace{0.3mm} \mathrm{d}\varphi \hspace{0.3mm}  \mathrm{d}h \bigg). 
\label{eq:nnf_lsp}
\end{align}
\end{figure*}

Combining Lemmas~\ref{lemma_ins_paradox} and \ref{hl_approx}, we note that the PDF of half-lengths of the sticks that pass through the typical vehicle in PLM can be approximated as $\tilde{f}_{H}(h) \approx h \hat{f}_{H}(h)/\mathbb{E}[H]$. Fig.~\ref{fig:fitted_lm} illustrates their histogram and the fitted density functions $\tilde{f}_{H}(h)$ and $f_{H}(h)$. We observe that $h \hat{f}_{H}(h)/\mathbb{E}[H]$ provides a good fit to the histogram, validating the approximation $\hat{f}_{H}(h)$.  

To appreciate the differences between the PSP-PPP and PLM-PPP, we consider the case where the half-lengths in the PSP are Rayleigh distributed as in the PLM. Let $N_{o}(r)$ denote the number of neighbors to the typical vehicle at $o$ within a distance $r$. Fig.~\ref{fig:lsp_lm_var_mean} compares the mean and variance of $N_o(r)$ in the PLM-PPP and PSP-PPP. The correlation among the stick lengths resulting from the touch-and-stop mechanism in the PLM-PPP leads to a smaller variance of $N_{o}(r)$ than in the PSP-PPP, where the stick lengths are independent. On the other hand, $\mathbb{E}[N_o]$ is the same for both the PSP-PPP and PLM-PPP when both follow the same distribution for half-lengths as their first-order statistics are the same. Also, the statistics of $N_{o}$ for the PLM-PPP and PSP-PPP differ significantly from that for a 2D PPP of equivalent intensity, highlighting the differences in having the street geometry and not.
\begin{figure}%
\centering
\subfloat[]{\includegraphics[scale=0.6]{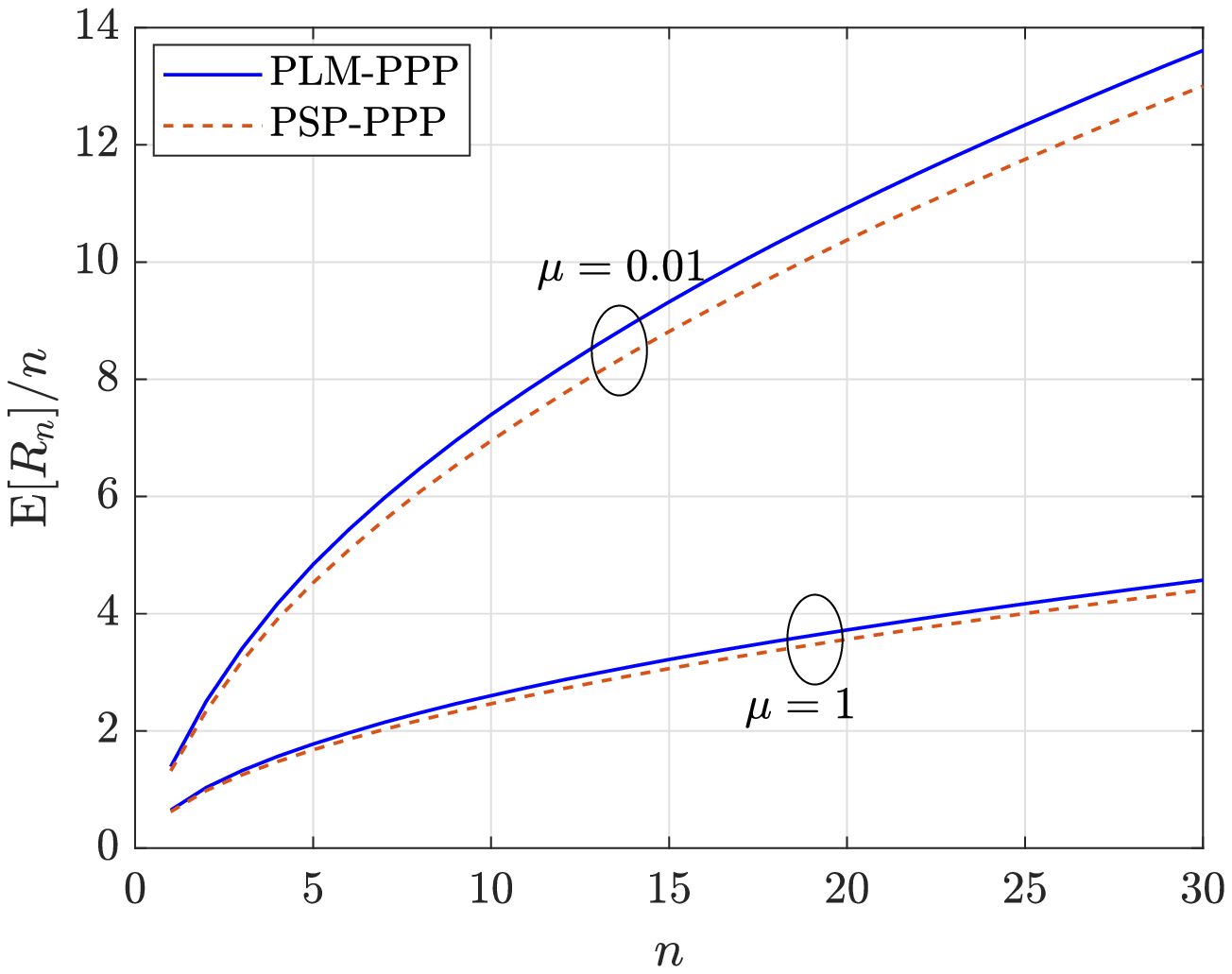}\label{fig:lsp_lm_nn}}\hfill
\subfloat[]{\includegraphics[scale=0.6]{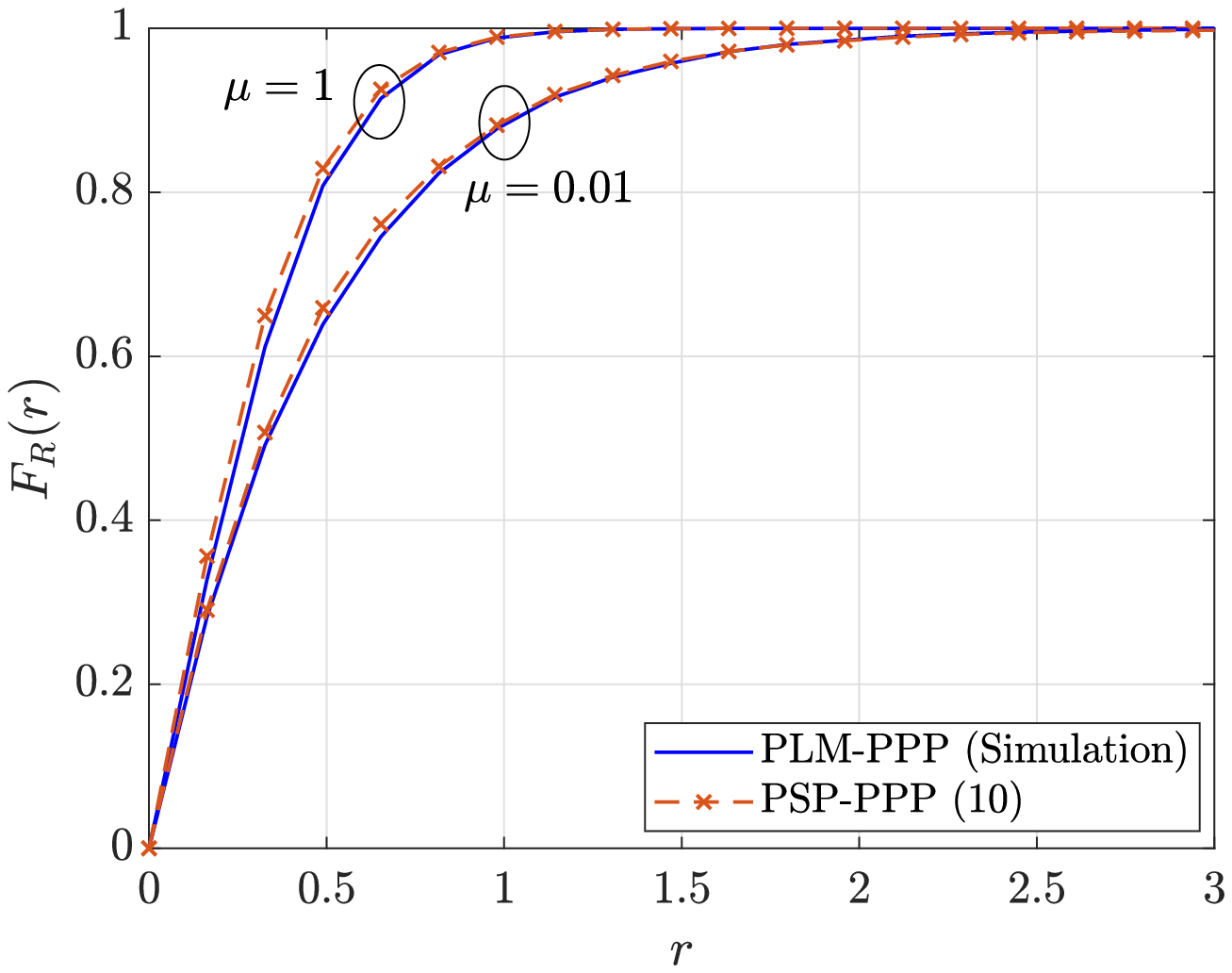}\label{fig:lsp_lm_nn1}}
\caption{Comparison of (a) mean distance from the typical general vehicle to its $n$th-nearest neighbor and (b) nearest-neighbor distance distributions in the PLM-PPP and PSP-PPP given by~\eqref{eq:nnf_lsp} with $f_{H}(h) = \hat{f}_{H}(h) = 2bh \exp(-bh^{2})$, where $b = 1.04$ for $\mu = 1$, and $0.0103$ for $\mu = 0.01$. $\lambda = 0.3$. }
\end{figure}

\setcounter{equation}{8}
\begin{lemma}
\label{nnf}
The nearest-neighbor distance distribution for a vehicular network with the street system characterized by Definition~\ref{street_system} can be decomposed as
\begin{align}
F_{R}(r) &= 1-(1-F_{R,\mathcal{V}_{o}^{m}}(r))(1-F_{R,\mathcal{V}_{!}}(r)),
\label{eq:nnf}
\end{align}
where $F_{R,\mathcal{V}_{o}^{m}}(r)$ is the probability of finding a neighbor in $\mathcal{V}_{o}^{m}$ within distance $r$, and $F_{R,\mathcal{V}_{!}}(r)$ is with respect to $\mathcal{V}_{!}$. $F_{R,\mathcal{V}_{!}}(r)$ also denotes the contact distance distribution, the distribution of the distance from an arbitrary location in the plane to the nearest vehicle in $\mathcal{V}$.
\end{lemma}
\begin{IEEEproof}
See Appendix~\ref{appendix:nnf_decomp_proof}.
\end{IEEEproof}

\begin{lemma}
\label{nnf_og}
The nearest-neighbor distance distribution for the OG-PPP/PLP-PPP is $F_{R}(r) = 1-\exp (-m \lambda  r - 2 \mu \smallint_{0}^{r} (1-\exp(-2 \lambda \sqrt{r^2-u^2}\hspace{0.3mm})) \mathrm{d}u )$, 
where $m \in \lbrace{2,4 \rbrace}$.
\end{lemma}
\begin{IEEEproof}
See Appendix~\ref{appendix:nnf_og_proof}.
\end{IEEEproof}

An alternative proof of the nearest-neighbor distance distribution for the case $m =2$ in the PLP-PPP can be found in~\cite{choi}. { Lemma~\ref{nnf_og} presents a slightly more general result that also applies to an intersection vehicle and the OG-PPP, which can be obtained by rotating each line $L \in \Xi_{L}$ (Definition~\ref{def:line_system}) constituting the PLP-PPP such that they are orthogonal to each other.}

\begin{theorem}
\label{th_nnf_lsp}
The nearest-neighbor distance distribution for the PSP-PPP is given by~\eqref{eq:nnf_lsp}, 
where $\ell (\gamma, \phi,\varphi)= \ell_1  (\gamma, \phi,\varphi) 
\mathds{1}_{\gamma \leq r} +  \ell_2(\gamma, \phi,\varphi)  \mathds{1}_{\gamma > r}$. $\ell_{1}(\gamma, \phi,\varphi), \ell_{2}(\gamma, \phi,\varphi) = \vert\min(u_{1}, h) \pm \min(u_{2}, h) \vert$, where $
u_{1}, u_{2}  = \vert -\gamma \cos(\phi-\varphi) \pm \sqrt{r^2-\gamma^{2}\sin^{2}(\phi-\varphi)} \vert$. $\tilde{f}_{H}(h) = h {f}_{H}(h)/\mathbb{E}[H]$, and $m \in \lbrace{2,4 \rbrace}$.
\end{theorem}
\begin{IEEEproof}
See Appendix~\ref{appendix:nnf_lsp}.
\end{IEEEproof}

Fig.~\ref{fig:lsp_lm_nn} shows the normalized mean distance $\mathbb{E}[R_{n}]/n$ to the $n$th-nearest neighbor in the PLM-PPP and PSP-PPP with Rayleigh distributed half-lengths. The rate of change in $\mathbb{E}[R_{n}]/n$ of the PLM-PPP from that of the PSP-PPP is at most 6\% for $\mu = 0.01$, and 5\% for $\mu = 1$. Fig.~\ref{fig:lsp_lm_nn1} compares the nearest-neighbor distance distributions for different values of $\mu$. We see that the nearest-neighbor distance distribution in the PLM-PPP is tightly upper bounded by that in the PSP-PPP in accordance with Fig.~\ref{fig:lsp_lm_nn}, which shows that the mean distance to the nearest neighbor in the PLM-PPP is tightly lower bounded by that in the PSP-PPP. We presume that the inference obtained from Fig.~\ref{fig:lsp_lm_nn1} extends to the $n$th-nearest neighbor for $n > 1$ as well.

\setcounter{equation}{15}
\begin{figure*}[b]
\hrule
\begin{align}
&\mathcal{L}_{I_{o}^{m}}^{\mathrm{PSP-PPP}}(s)  = \bigg(\int \limits_{0}^{\infty} \bigg( \frac{1}{2h} \int \limits_{-h}^{h} \exp \bigg(\hspace{-1mm}- \lambda p  s^{\delta/2}  \int \limits_{(-w-h) s^{-\delta/2}}^{(-w+h) s^{-\delta/2}} \frac{1}{ 1 + v^{{2/\delta}}} \mathrm{d}v\bigg) \mathrm{d}w \bigg)\tilde{f}_{H}(h) \hspace{0.3mm} \mathrm{d}h \bigg)^{m/2}. \label{eq:lio_psp} \\
&\mathcal{L}_{I_{!}}^{\mathrm{PSP-PPP}}(s)  = \exp\bigg(\hspace{-1mm}-\frac{\mu}{\pi} \int \limits_{0}^{\infty} \int \limits_{0}^{\pi} \int \limits_{0}^{2 \pi} \int \limits_{0}^{\infty} \bigg(1 - \exp\bigg(\hspace{-1mm}-\lambda p \int \limits_{-h}^{h} \bigg({1+ \bigg(\frac{\mathfrak{F}(\gamma, u, \phi, \varphi)}{s^\delta}\bigg)^{1/\delta}}\bigg)^{-1} \mathrm{d}u \bigg)\bigg) \gamma {f}_{H}(h)  \hspace{0.3mm}\mathrm{d}\gamma \hspace{0.3mm} \mathrm{d}\phi \hspace{0.3mm} \mathrm{d}\varphi \hspace{0.3mm} \mathrm{d}h \bigg).
\label{eq:lir_psp}\\
\setcounter{equation}{18}
&p_{2}^\mathrm{PLM-PPP} \approx {p}_{2}^\mathrm{PSP-PPP} = \mathcal{L}_{I_{o}^{2}}^{\mathrm{PSP-PPP}}(\theta D^{\alpha})\mathcal{L}_{I_{!}}^{\mathrm{PSP-PPP}}(\theta D^{\alpha}).
\label{eq:ps_lm_gen} \\
&p_{3}^\mathrm{PLM-PPP} \approx  {p}_{2}^\mathrm{PSP-PPP}  \times \int \limits_{0}^{\infty}\exp \bigg(\hspace{-1mm}- \lambda p D \theta^{\delta/2} \int \limits_{0}^{\frac{2h}{D \theta^{\delta/2}}} \frac{1}{ 1 + v^{{2/\delta}}} \mathrm{d}v\bigg)\hat{f}_{H}(h) \hspace{0.3mm}  \mathrm{d}h.
\label{eq:ps_lm_tjn}
\end{align}
\end{figure*}
\begin{conjecture}
\label{conj_lsp_lm_nn}
The distance from the typical general vehicle to the $n$th-nearest neighbor in the PLM-PPP stochastically dominates that distance in the PSP-PPP with Rayleigh distributed half-lengths.
\end{conjecture}

To facilitate the comparison of the success probabilities in the general street-based Cox models and the homogeneous PPP, we recall the success probability of the typical vehicle in a PPP. Let $\Phi_{d}$ denote a stationary $d-$dimensional PPP of intensity $\lambda_{d}$ and $c_{d}$ denote the volume of a unit $d-$dimensional ball. In particular, $c_{1} = 2$ and $c_{2} = \pi$. 

\setcounter{equation}{10}
\begin{lemma}[{{{\cite{haenggi}, Sec. 5.2}}}]
\label{lemma_nd_ppp}
The success probability $p_{\mathrm{s}}$ of the typical vehicle in $\Phi_{d}$ is 
\begin{align}
p^{\Phi_{d}}_{\mathrm{s}} = \exp(-  c_{d} \lambda_{d}  D^{d} \theta^{\delta} \Gamma(1+\delta') \Gamma(1-\delta') ),
\label{eq:ps_nd_ppp}
\end{align}
where $\delta' = d/\alpha$.
\end{lemma}
Next, we analyze the success probabilities of the typical vehicle in Cox vehicular networks.

\section{Success Probabilities}
Using~\eqref{eq:sir_expr} and the notations in Section~\ref{sec:notation}, we express the success probability $p_{m}$ of the typical vehicle of order $m$ as
\begin{align}
p_{m} & = \mathbb{P}(g > \theta D^{\alpha} I)  \nonumber \\
& = \mathbb{E}[\exp(-\theta D^{\alpha} (I_{o}^{m} + I_{!})]. \label{eq:sir_joint}
\end{align}
We can simplify~\eqref{eq:sir_joint} for all the Cox vehicular networks considered but the PLM-PPP as
\begin{align}
p_{m} & \stackrel{(a)}= \mathbb{E}[\exp(-\theta D^{\alpha} I_{o}^{m})] \mathbb{E}[\exp(-\theta D^{\alpha} I_{!})] \label{eq:sir_indep} \\
& \stackrel{(b)} = \mathcal{L}_{I_{o}^{m}}(s)\mathcal{L}_{I_{!}}(s)\vert_{s= \theta D^{\alpha}}, \label{eq:sir}
\end{align}
where $(a)$ is due to the independence of the PPPs on the streets, and $(b)$ applies the definition of the Laplace transform. For the PLM-PPP, \eqref{eq:sir_indep} does not hold as the length of each street is dependent on that of other streets.

\begin{figure*}
\centering
\subfloat[]{\includegraphics[scale = 0.4] {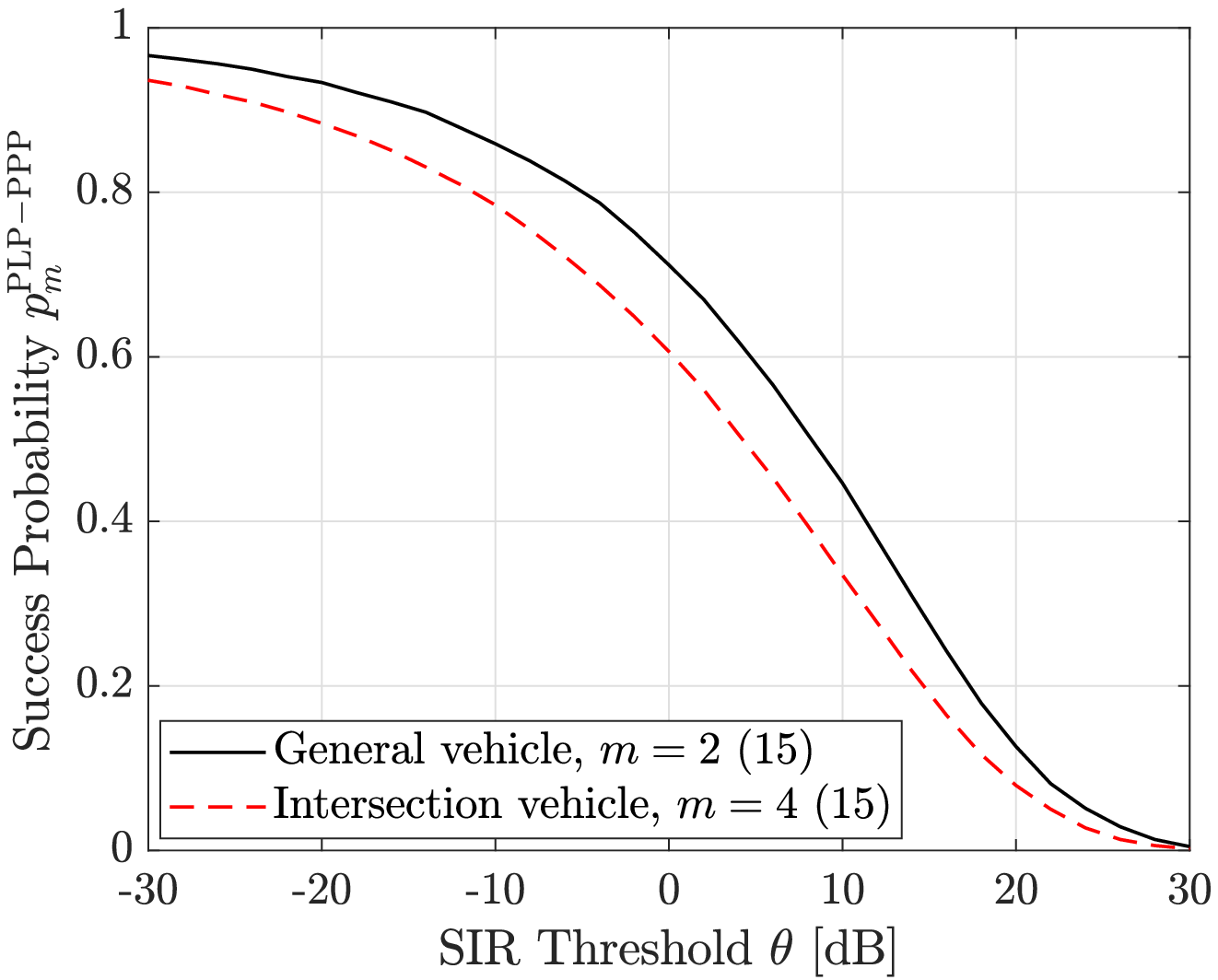}
 } \hfill
\subfloat[]{\includegraphics[scale = 0.4]{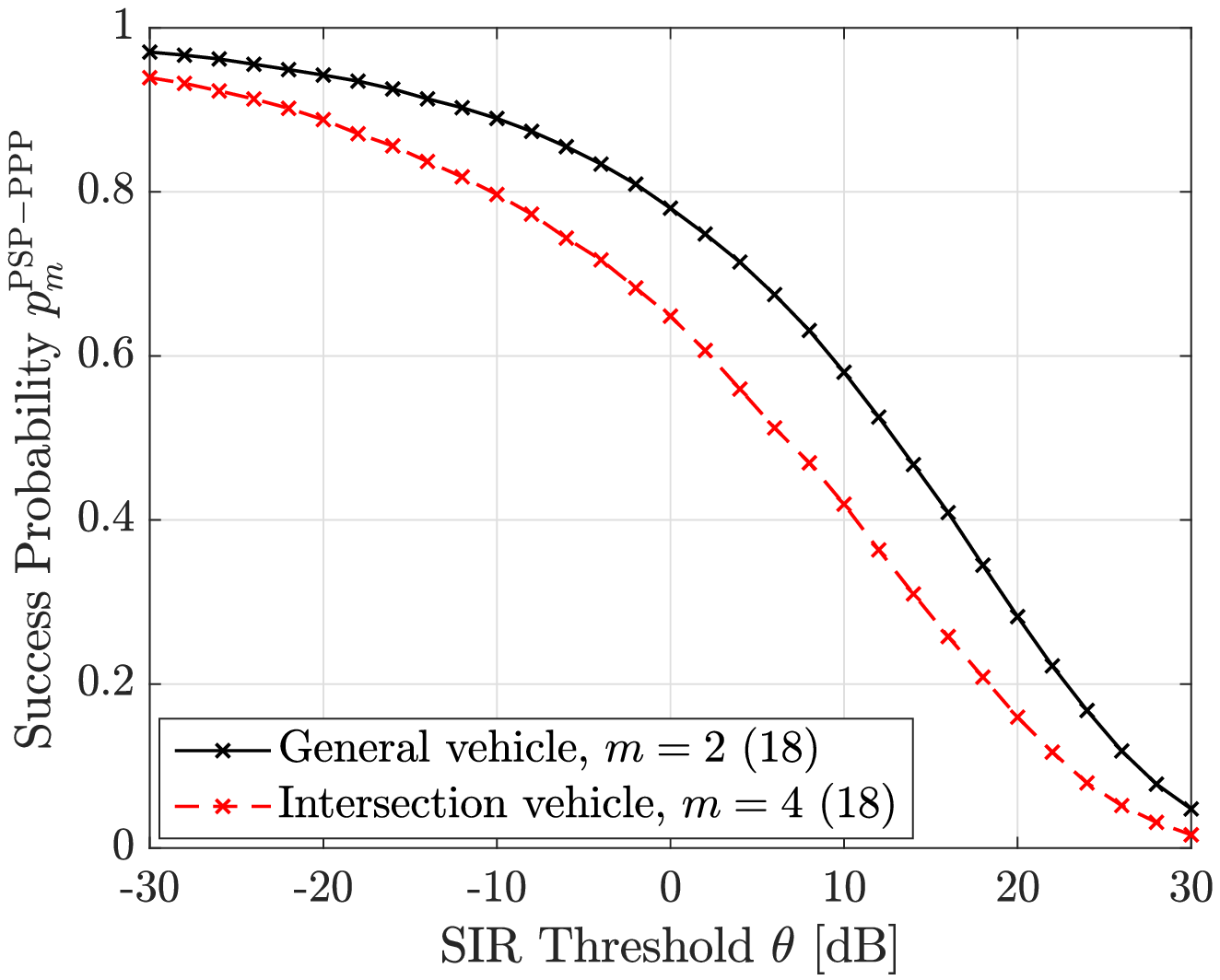}
 }\hfill
 \subfloat[]{\includegraphics[scale = 0.4]{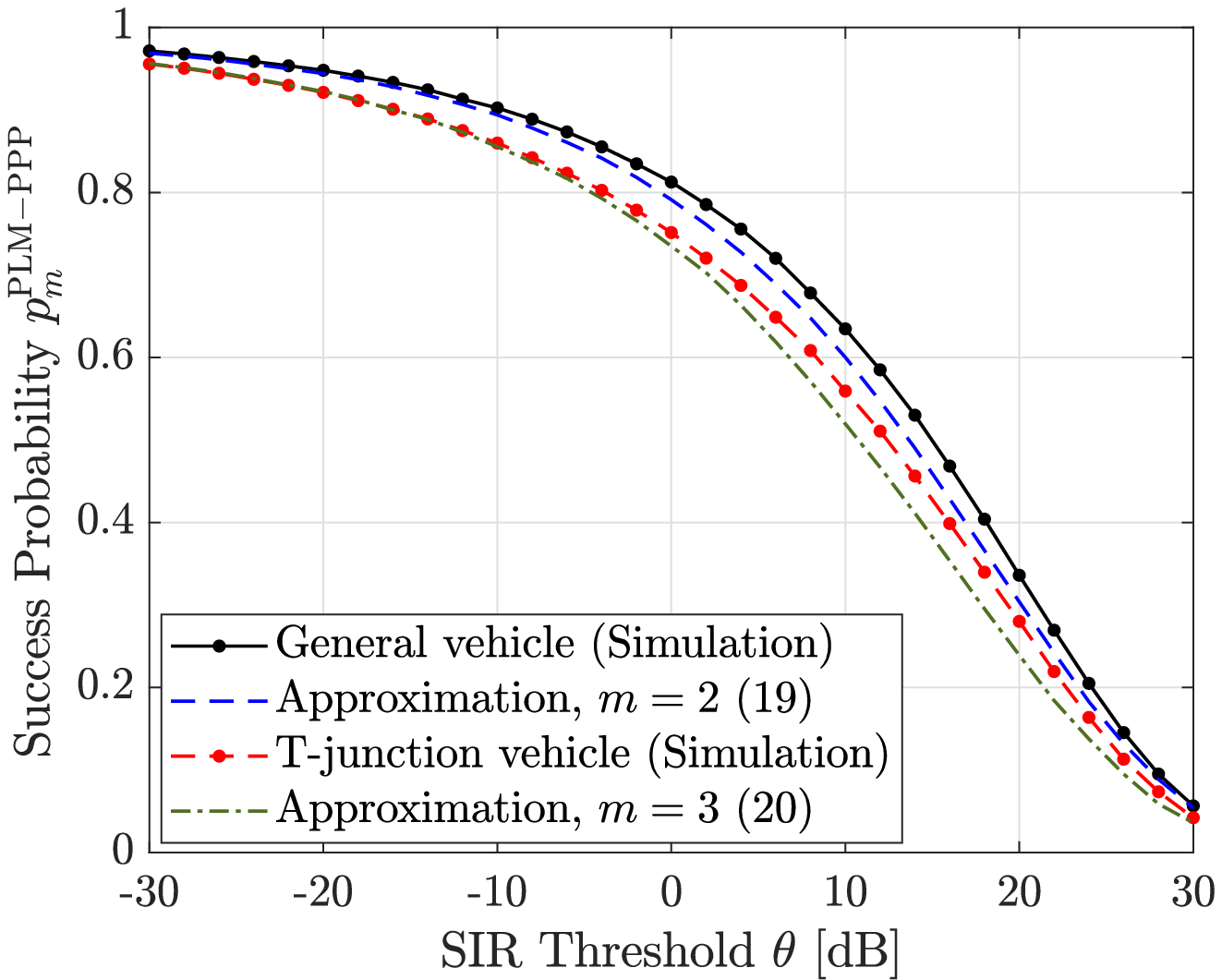}{\label{fig:plm_comp}}
}\hfill
\caption{\label{fig:ps_gen_int} {Success probabilities of the typical general and intersection/T-junction vehicles in the (a) PLP-PPP (b) PSP-PPP and (c) PLM-PPP. $\lambda p = 0.3, D = 0.25$, $\alpha = 4$, $\mu = 2$, $0.1$, and $0.3$ for the PLP, PSP, and PLM, respectively. $f^{\mathrm{PSP}}_{H}(h) = \delta(h-10)$. The equation numbers are given in parentheses in the legends.}}
\end{figure*}

\begin{proposition}
\label{prop_ps_og_plp}
The success probability of the typical vehicle in the OG-PPP/PLP-PPP is 
\begin{align}
 p_{m}  = \exp\bigg(\hspace{-1mm}&- m \lambda p D  \theta^{\delta/2} \Gamma(1+\delta/2)\Gamma(1-\delta/2) \nonumber \\
 & - 2 \mu \int_{0}^{\infty} (1- \mathcal{L}_{I_x}(\theta D^{\alpha})) \hspace{0.3mm} \mathrm{d}x\bigg), \label{eq:ps_og_ppp}
 \end{align}
 where $
\mathcal{L}_{I_x}(s) = \exp \big(-  \lambda p  s^{\delta/2} \int_{v_x}^{\infty} \frac{1}{\left( 1 + v^{{1/\delta}}\right) \sqrt{v - v_x}} \mathrm{d}v\big)$ with $v_x = \frac{x^2}{s^{\delta}}$, and $m \in \lbrace 2, 4 \rbrace$.
\end{proposition}
\begin{IEEEproof}
See Appendix~\ref{appendix:plp}.
\end{IEEEproof}
The success probability of the typical vehicle of order 2 in the PLP-PPP is derived in~\cite{chetlur}. The success probability~\eqref{eq:ps_og_ppp} depends only on the distances of the interferers to the typical vehicle, not their orientations. In~Appendix \ref{appendix:plp}, we give a general proof that shows the effect of the order of the vehicle and the irrelevance of the orientations on the success probability.

\begin{proposition}
\label{prop_lio_psp}
The Laplace transform of the interference from the vehicles on streets that pass through the typical vehicle of order $m \in \lbrace 2,4 \rbrace $ in the PSP-PPP with half-length density function $f_{H}(h)$ is given by~\eqref{eq:lio_psp}, where $\tilde{f}_{H}(h) = hf_{H}(h)/\mathbb{E}[H]$.
\end{proposition}
\begin{IEEEproof}
See Appendix~\ref{appendix:lio_psp}.
\end{IEEEproof}

\begin{proposition}
\label{prop_lir_psp}
The Laplace transform of the interference from the vehicles on all but the streets that pass through the typical vehicle in the PSP-PPP with half-length density function $f_{H}(h)$ is given by~\eqref{eq:lir_psp}, where
$\mathfrak{F}(\gamma, u, \phi, \varphi) = \gamma^2+u^2+2\gamma u \cos(\phi-\varphi)$.
\end{proposition}
\begin{IEEEproof}
See Appendix~\ref{appendix:lir_psp}.
\end{IEEEproof}
\setcounter{equation}{17}
Following~\eqref{eq:sir}, the success probability of the typical vehicle in the PSP-PPP is 
\begin{align}
p_{m}^\mathrm{PSP-PPP} = \mathcal{L}_{I_{o}^{m}}^{\mathrm{PSP-PPP}}(\theta D^{\alpha})\mathcal{L}_{I_{!}}^{\mathrm{PSP-PPP}}(\theta D^{\alpha}),
\end{align}
where $\mathcal{L}_{I_{o}^{m}}^\mathrm{PSP-PPP}(s)$ and $\mathcal{L}_{I_{!}}^{\mathrm{PSP-PPP}}(s)$ are given by \eqref{eq:lio_psp} and \eqref{eq:lir_psp}, respectively. 
\begin{proposition}
\label{lem_lp_ppp}
The success probabilities of the typical general vehicle (order 2) and the typical T-junction vehicle (order 3) in the PLM-PPP can be approximated as in~\eqref{eq:ps_lm_gen} and \eqref{eq:ps_lm_tjn}, respectively. 
$\mathcal{L}_{I_{o}^{2}}^\mathrm{PSP-PPP}(s)$ and $\mathcal{L}_{I_{!}}^{\mathrm{PSP-PPP}}(s)$ in~\eqref{eq:ps_lm_gen} are given by \eqref{eq:lio_psp} and \eqref{eq:lir_psp}, respectively, with $f_{H}(h) = \hat{f}_{H}(h) = 2bh\exp(-bh^{2})$.
\end{proposition}
\begin{IEEEproof}
See Appendix~\ref{appendix:lp}.
\end{IEEEproof}

Fig.~\ref{fig:ps_gen_int} compares the success probabilities of the typical general and intersection/T-junction vehicles. We omit the plot for the success probability of the typical vehicle in the OG-PPP as it is the same as that in the PLP-PPP by Proposition~\ref{prop_ps_og_plp}. We see that the success probability of the typical general vehicle is higher than that of the typical intersection/T-junction vehicle in all the Cox vehicular networks. The reason is that as two streets pass through or end at the typical vehicle at an intersection or a T-junction, the received interference is higher compared to the typical general vehicle through which only one street passes.

\begin{remark}
{ Fig.~\ref{fig:plm_comp} indicates that the success probability of the typical general vehicle in the PLM-PPP is tightly lower bounded by that in the PSP-PPP with Rayleigh distributed half-lengths.}
\end{remark}
\begin{remark}
The approximation~\eqref{eq:ps_lm_tjn} to the success probability of the typical T-junction vehicle serves as a tight lower bound {(see Fig.~\ref{fig:plm_comp})}. The integral term on the right-hand side of~\eqref{eq:ps_lm_tjn} is the Laplace transform of the interference from the vehicles on the street that has its endpoint at the origin. We can rephrase~\eqref{eq:ps_lm_tjn} as follows. The success probability of the typical T-junction vehicle in the PLM-PPP is tightly lower bounded by the success probability of the typical vehicle (with one street passing through) in the network formed by conditioning a street in the PSP-PPP (with Rayleigh distributed half-lengths) such that one of its endpoints is at the origin. 
\end{remark}

Next, we analyze the equivalence between the spatial models using their properties and the success probabilities of the typical vehicle in those models. 
\section{Equivalence of Spatial Models}
\label{equiv_bipolar}
To differentiate the street intensities of the models OG, PLP, PSP, and PLM, we denote them as $\mu_{{\mathrm{OG}}}$, $\mu_{\mathrm{PLP}}$, $\mu_{\mathrm{PSP}}$, and $\mu_{\mathrm{PLM}}$, respectively.
\subsection{OG-PPP and PLP-PPP}
\begin{theorem}
\label{th_og_plp}
The PLP-PPP of street intensity $\mu_{\mathrm{PLP}}$ is strictly equivalent to the OG-PPP with street intensity $\mu_{\mathrm{OG}} = \mu_{\mathrm{PLP}}$.
\end{theorem}
\begin{IEEEproof}
The equivalence is a direct consequence of the success probabilities of the typical vehicle in the OG-PPP and PLP-PPP given in Proposition 1.
\end{IEEEproof}

\subsection{PSP-PPP and PLP-PPP}
\begin{theorem}
\label{prop_lsp_plp}
Let $H \triangleq c H_{1}$, where $c$ is a constant and $H_{1}$ is a random variable with mean 1. The PLP is the limiting process of the PSP as $c \to \infty$ and $\mu_{{\mathrm{PSP}}} \to 0$ such that $2\mu_{{\mathrm{PSP}}} \mathbb{E}[H] = 2 c \mu_{{\mathrm{PSP}}} = \mu_{{\mathrm{PLP}}}$.
\end{theorem}

\begin{IEEEproof}
\begin{figure}
\centering
\includegraphics[scale=0.4]{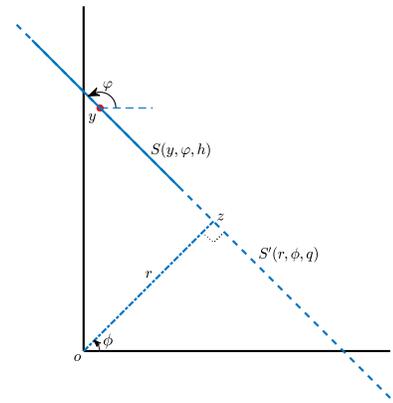}
\caption{Reparametrization of the PSP. The stick $S(y, \varphi, h) \in \Xi_{\mathrm{S}}$ is extended to form a line $S'(r, \phi,q)$. The perpendicular from the extended stick is at a distance $r$ from $o$ and forms an angle $\phi$ with the $x-$axis.}
\label{fig:lsp_plp}
\end{figure}
{ We formalize the heuristic arguments on the relation between the PLP and PSP given in \cite{parker}. From Definitions~\ref{def:line_system} and \ref{def:stick_system}, we learn that the parameter spaces of line and stick are different. To compare the line process with the stick process, we first establish compatible parametrizations. Let $S'(r,\phi, q)$ denote the infinitely extended stick $S(y,\varphi,h)$, where
$q = \overline{yz}$ is the distance between the midpoint of the stick $y = (u,v)$ and $z = (r \cos\phi, r\sin\phi)$, the closest point to the origin from $S'(r,\phi, q)$. Fig.~\ref{fig:lsp_plp} illustrates $S'(r,\phi, q)$ and $S(y,\varphi,h)$ using overlaid dashed and solid lines, respectively. 
We can express $y =(u,v)$ and $\varphi$ in terms of $(r, \phi)$ and $q$ as
$u = r\cos\phi-q \sin\phi, \hspace{0.3mm} 
v = r\sin\phi+q \cos\phi, \hspace{0.3mm} \text{and} \hspace{1mm} \varphi = \phi-\pi/2$.
The differential element $\mathrm{d}u \hspace{0.3mm}\mathrm{d}v \hspace{0.3mm} \mathrm{d}\varphi$ equals $\mathrm{d}r \hspace{0.3mm} \mathrm{d}\phi \hspace{0.3mm} \mathrm{d}q$, {\em{i.e.,}} the alternate parametrizations are equivalent. 

By~\eqref{eq:line}, the line is just the projection of the stick from the parameter space $(r, \phi, q)$ to $(r, \phi)$ when the latter is extended to infinity. However, we learn from Lemmas~\ref{lemma_2d_og} and \ref{lemma_lsp_2d} that the properties of the PLP and PSP differ. Then, to obtain the PLP from PSP, we need to equate their statistical properties. Equating the mean total street length per unit area in the PLP and PSP given in Lemmas~\ref{lemma_2d_og} and \ref{lemma_lsp_2d}, we obtain $2 \mu_{{\mathrm{PSP}}}\mathbb{E}[H] = \mu_{{\mathrm{PLP}}}$. For $\mu_{{\mathrm{PSP}}}\mathbb{E}[H] = c \mu_{{\mathrm{PSP}}}$ to remain finite, $\mu_{{\mathrm{PSP}}}$ should go to zero as $c \to \infty$.}
\end{IEEEproof}

\begin{corollary}
The PSP-PPP, and its limiting case, the PLP-PPP, are strictly equivalent. As the PLP and PSP are identically distributed as $c \to \infty$ and $\mu_{\mathrm{PSP}} \to 0$, equivalence is not restricted to PPPs on the streets but also holds for general point processes of vehicles.
\end{corollary}

\subsection{PLM-PPP and PSP-PPP}
\label{section: plm_psp_equi}
\begin{figure}
\centering
{\includegraphics[scale=0.585]{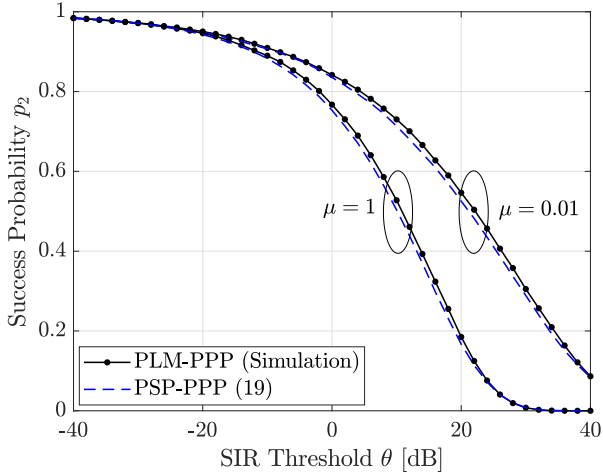}}
\caption{Success probability of the typical general vehicle in the PLM-PPP vs. that in the PSP-PPP-based approximation given by $(19)$ with $f_{H}(h) = 2bh\exp(-bh^2)$, where $b = 1.04$ for $\mu = 1$, and $0.0103$ for $\mu = 0.01$. $\lambda p = 0.3$, $D=0.25$, and $\alpha = 4$.}
\label{fig:lm_equi}
\end{figure}

We learned from Proposition 4 that the success probability of the typical general vehicle in the PLM-PPP is approximated by that in the PSP-PPP with the same half-length density function as the PLM-PPP. Furthermore, from Figs.~\ref{fig:lsp_lm_nn} and \ref{fig:lsp_lm_nn1}, we infer that the $n$th-nearest-neighbor distance distribution in the PLM-PPP is tightly upper bounded by that in the PSP-PPP. Consequently, we can deduce that the approximation to the success probability in the PLM-PPP is tight. Fig.~\ref{fig:lm_equi} compares the simulated success probability of the typical general vehicle in the PLM-PPP with its lower bound. The maximum difference between the success probabilities in the PSP-PPP and PLM-PPP is $\epsilon = 0.0297$ for $\mu = 0.01$ and 0.0219 for $\mu = 1$. Though the PLM can characterize T-junctions, it is too complex to permit an exact analytical expression.  
\begin{remark}
The PLM-PPP is $\epsilon-$equivalent with $\epsilon \ll 1$ to the PSP-PPP with the same half-length distribution and street intensity as the PLM-PPP. Consequently, the PSP-PPP serves as a good substitute for the PLM-PPP.
\end{remark}
We see from Fig.~\ref{fig:lm_equi} that the success probabilities of the typical general vehicle in the PLM-PPP and PSP-PPP with Rayleigh distributed half-lengths are even closer in the asymptotic regions of $\theta$ than in the middle regions of $\theta$. Theorem~\ref{th_psp_asym} proves this observation formally. 
\begin{theorem}
\label{th_psp_asym}
The PLM-PPP is asymptotically equivalent in both the lower and upper regimes of $\theta$ to the PSP-PPP with the same half-length density as the PLM-PPP. 
\end{theorem}
\begin{IEEEproof}
{ First, we study the asymptotic behavior of the PSP-PPP as $\theta \to 0$ and $\infty$. Then, we compare them with that of the PLM-PPP.
\setcounter{equation}{20}
\begin{lemma}
\label{th_lsp_zero}
As $\theta \to 0$, the PSP-PPP behaves like a vehicular point process formed only by the typical vehicle's streets, i.e.,
\begin{align}
1-p_{m}^{\mathrm{PSP-PPP}}(\theta) = \Theta(\theta^{\delta m/4}).
\end{align}
\end{lemma}
\begin{IEEEproof}
See Appendix~\ref{appendix:th_lsp_zero}. 
\end{IEEEproof}
\begin{lemma}
\label{th_lsp_infty}
As $\theta \to \infty$, the success probability of the typical vehicle in the PSP-PPP tends to that in a 2D PPP, i.e.,
\begin{equation}
p_{m}^\mathrm{PSP-PPP}(\theta) \sim \exp(-  \pi \lambda_{2} p D^{2} \theta^{\delta}\Gamma(1+\delta)\Gamma(1-\delta)),  \label{eq:asym_infty_lsp}
\end{equation}
where $\lambda_{2}  = 2 \mu \lambda  \mathbb{E}[H] $ is the 2D intensity of the PSP-PPP.
\end{lemma}
\begin{IEEEproof}
See Appendix~\ref{appendix:th_lsp_infty}.
\end{IEEEproof}

Equipped with the two lemmas, we continue with the proof of the theorem. As $\theta \to 0$, for SIR $> \theta$ to hold, it suffices not to have any interferers within a small disk around the typical vehicle. With high probability, the small disk intersects only the street(s) passing through the typical vehicle. Consequently, the system performance converges to that of only the typical vehicle's streets as $\theta \to 0$. The outage probability of the typical vehicle due to its streets alone is proportional to $\lambda p \theta^{\delta m/4}$ as $\theta \to 0$. On the other hand, as $\theta \to \infty$, for SIR $> \theta$, a large disk around the typical vehicle must be devoid of interferers. It follows from Lemma~\ref{th_lsp_infty} that the street geometry outside a large disk does not matter as $\theta \to \infty$ since the PSP-PPP is similar to a 2D PPP at a large scale.

Lemma~\ref{th_lsp_zero} extends to the PLM-PPP as the interaction between the sticks is irrelevant when $\theta \to 0$. Also, the success probability of the typical general vehicle with respect to its street alone in the PLM-PPP is the same as that in the PSP-PPP with the same half-length density as the PLM-PPP. In Appendix~\ref{appendix:th_lsp_infty}, we reasoned that the PSP-PPP behaves like a 2D PPP as $\theta \to \infty$ through mapping all the points on the sticks to their respective midpoints. The same logic holds for the PLM-PPP as it is also formed by sticks whose midpoints form a PPP. As the PLM-PPP and PSP-PPP with the same half-length density as the PLM-PPP behave like 2D PPPs of the same intensities as $\theta \to \infty$, they are equivalent as $\theta \to \infty$. }
\end{IEEEproof}

\begin{figure*}
\centering
\subfloat[]{\includegraphics[scale = 0.4]{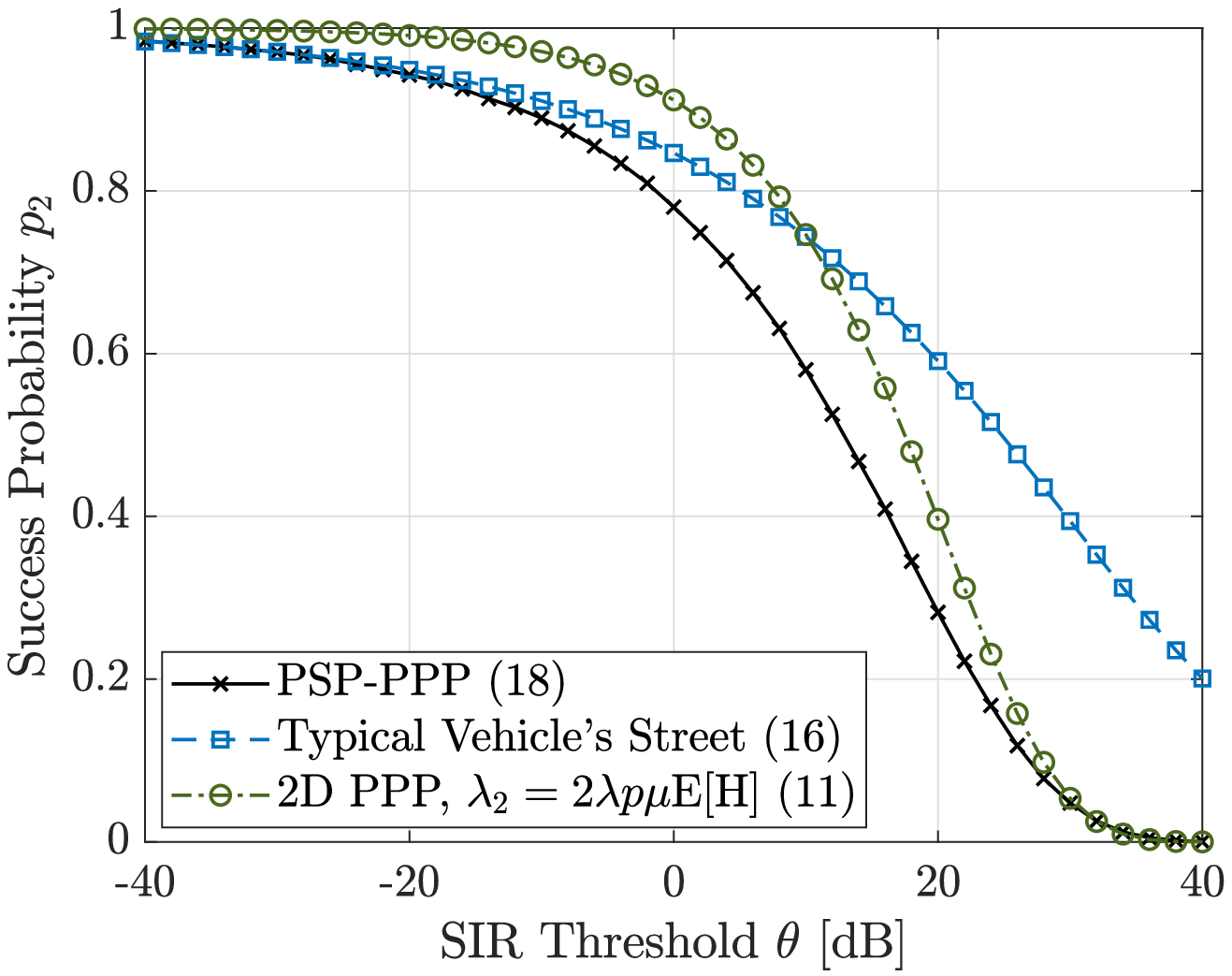}
 }\hfill
\subfloat[]{\includegraphics[scale = 0.4]{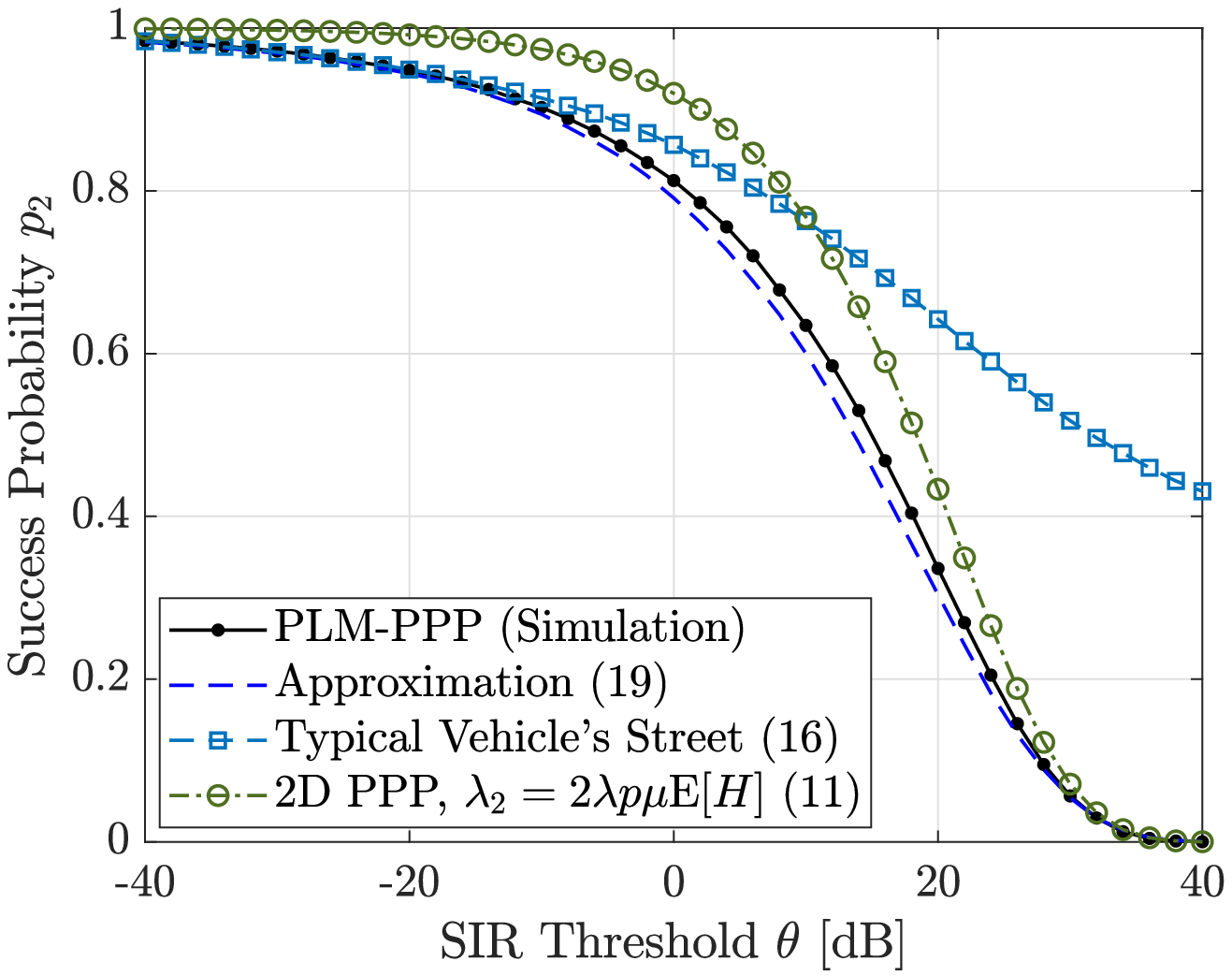}
} \hfill
\subfloat[]{\includegraphics[scale = 0.4] {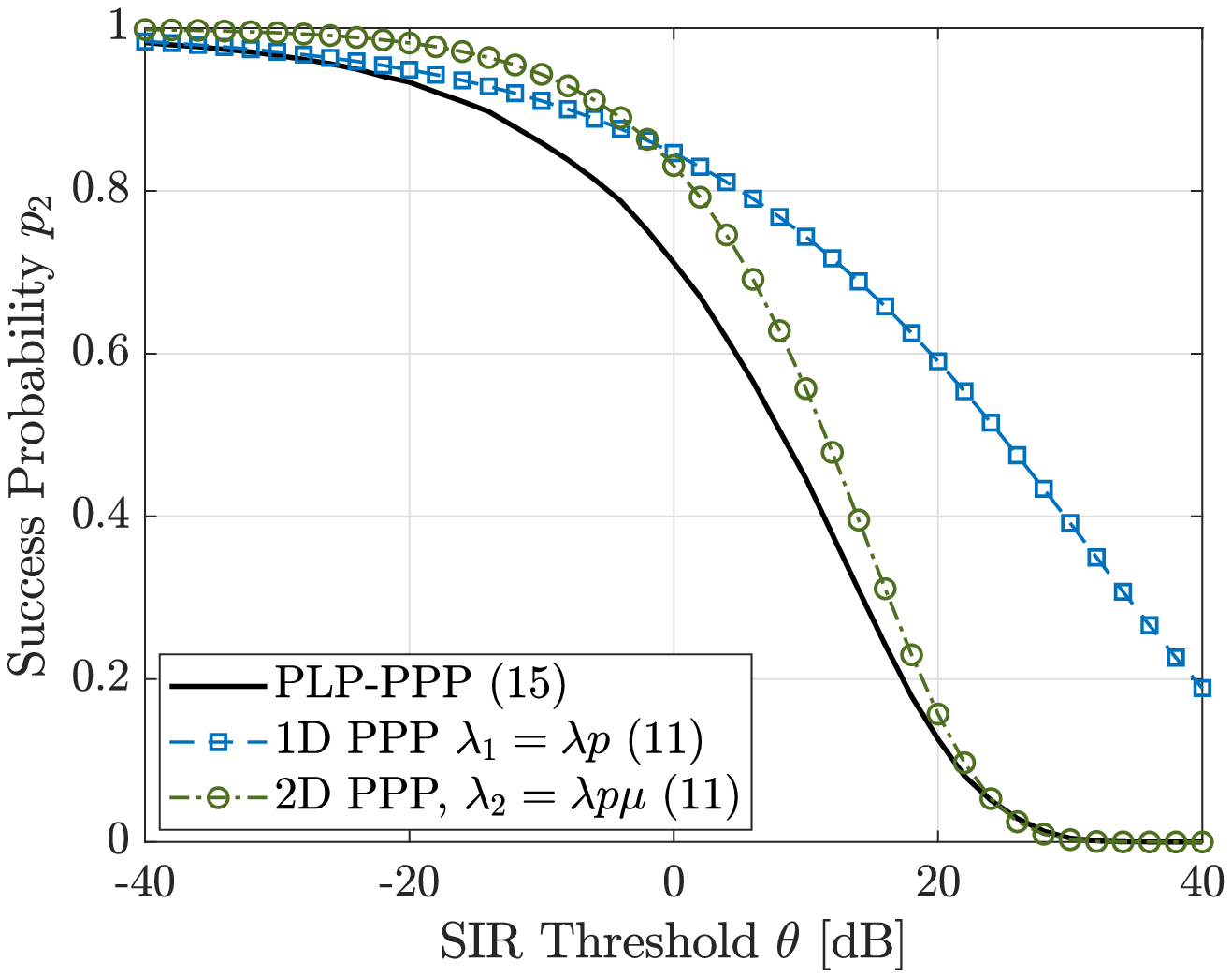}
 } \hfill
\caption{\label{fig:bounds_ps} {Success probability of the typical general vehicle in the (a) PSP-PPP (b) PLM-PPP (c) PLP-PPP vs. that of the typical vehicle in 1D and 2D PPPs. $\lambda p = 0.3, D = 0.25$, $\alpha = 4$, $\mu =  0.1$, $0.3$, and $2$ for the PSP, PLM, and PLP, respectively. $f^{\mathrm{PSP}}_{H}(h) = \delta(h-10)$. The equation numbers are given in parentheses in the legends.}}
\end{figure*}

\begin{table*}
\caption{Conditions for equivalence between the spatial models. }
\begin{center}
    \begin{tabular}{ | c | c | c |}
    \hline
    Model A & Model B & Conditions for Equivalence \\ \hline
    OG-PPP & PLP-PPP & $\mu_{\mathrm{OG}} =  \mu_{\mathrm{PLP}}$  \\ \hline
    PLP-PPP & PSP-PPP & $\mu_{{\mathrm{PSP}}} \to 0$ and $c \to \infty$ s.t.
    $2\mu_{{\mathrm{PSP}}} \mathbb{E}[H] = 2 c \mu_{{\mathrm{PSP}}} =  \mu_{{\mathrm{PLP}}}$ \\ \hline
    PSP-PPP & PLM-PPP & $\mu_{\mathrm{PSP}} = \mu_{\mathrm{PLM}}$, $f_{H}^{\mathrm{PSP}}(h) = f_{H}^{\mathrm{PLM}}(h)$ \\ \hline 
    \end{tabular}    
\end{center} 
\label{tab:equiv}
\end{table*}
Fig.~\ref{fig:bounds_ps} compares the success probabilities of the typical general vehicle in the street-based Cox vehicular networks with that of the typical vehicle in 1D and 2D PPPs. The success probability of the typical vehicle in the PSP-PPP and PLM-PPP tends to that in the network formed only by the typical vehicle's street(s) as $\theta \to 0$ and to that of the 2D PPP as $\theta \to \infty$, as given in Theorem~\ref{th_psp_asym} and the lemmas therein. The same holds for the PLP-PPP, as established in~\cite[Th. 4 and 5]{jeyaj_tppp}. As vehicles on each street in the PLP-PPP form a 1D PPP, the PLP-PPP behaves like a 1D PPP as $\theta \to 0$. 

Until now, we have assumed that the link distance $D$ is fixed. Next, we discuss the equivalence between the spatial models when the link distances are random. Here, the success probability is obtained by averaging the Laplace transform of the interference over the link distances.
\subsection{Equivalence Under Random Link Distances}
\label{equiv_nn}
\subsubsection{PLP-PPP vs. OG-PPP} 
The OG-PPP is just a rotational variant of the PLP-PPP. Both of them have the same statistical properties such as the mean total street length per unit area (Lemma~\ref{lemma_2d_og}), distribution of the distance to the nearest-neighbor (Lemma~\ref{nnf_og}), and the Laplace transform of the interference (Proposition~\ref{prop_ps_og_plp}). Hence, the OG-PPP and PLP-PPP are strictly equivalent even if the link distances are random. 

\subsubsection{PSP-PPP vs. PLP-PPP}
The PLP is the limiting process of the PSP as the lengths of the sticks extend to infinity and street intensity tends to zero (Theorem~\ref{prop_lsp_plp}). By the inherent nature of the PSP, it is equivalent to the PLP irrespective of the mode of communication.

\subsubsection{PLM-PPP vs. PSP-PPP}
\begin{figure}[]
\centering
\includegraphics[scale = 0.6]{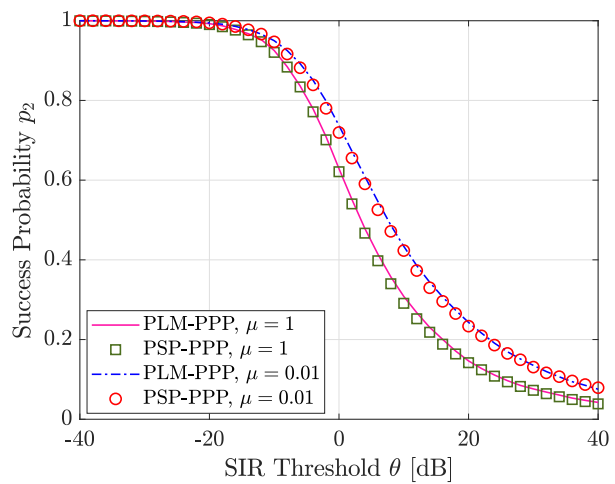}
\label{fig:lsp_lm_nrt}
\caption{Success probabilities of the typical general vehicle that receives a message from its nearest neighbor. $\lambda p = 0.5$ and $\alpha = 4$. $f_{H}(h) = \hat{f}_{H}(h) = 2bh\exp(-bh^2)$, where $b = 1.04$ for $\mu = 1$, and $0.0103$ for $\mu = 0.01$. }
\label{fig:lsp_lm_comp_nnc}
\end{figure}

We deduced from Figs.~\ref{fig:lsp_lm_nn} and \ref{fig:lsp_lm_nn1} that the $n$th-nearest-neighbor (or interferer) distance distribution in the PLM-PPP is tightly upper bounded by that in the PSP-PPP with Rayleigh distributed half-lengths. It follows that irrespective of the distribution of the distance to the intended transmitter, the PLM-PPP and PSP-PPP with Rayleigh distributed half-lengths are $\epsilon-$equivalent. Fig.~\ref{fig:lsp_lm_comp_nnc} validates the above inference for the case where the typical general vehicle receives a message from its nearest neighbor (transmitter) through simulations.  

\begin{remark}
The notion of equivalence enables us to consider only a representative set of spatial models to obtain insights on the effect of the network parameters, thereby reducing the computational costs and time associated with large-scale experiments and system-level simulations. The success probabilities of the typical vehicle in the OG-PPP, PLP-PPP, and PLM-PPP can be obtained from that in the PSP-PPP by a suitable mapping between the parameters as summarized in Table~\ref{tab:equiv}.
\end{remark} 

\section{Conclusions and Future Work}                                            
We developed a Coxian framework for the modeling and analysis of vehicular networks. The spatial models in this framework can characterize different street geometries involving intersections and T-junctions, and street lengths that can be independent or dependent on each other. Streets of infinite lengths and different orientations forming intersections can be characterized by PLPs and their rotational variants. PSPs and PLMs can model streets of varying finite lengths forming intersections and T-junctions, respectively. We evaluated the reliability of a vehicle-to-vehicle link in the Cox vehicular networks when the receiving vehicle is at an intersection, a T-junction, or a general location, and its transmitter is at a certain fixed distance. Our approach to defining the street system as a union of points of different orders facilitates general analytical results for different types of typical vehicles. The expressions for the reliability can be used to investigate the interplay among the network parameters such as data rate, street intensity, vehicle density, and the type of vehicle. 

The concept of equivalence demonstrates that one does not need different spatial models to analyze the reliability of a vehicle-to-vehicle link in different geographical regions. The models considered in the Coxian framework are equivalent in terms of reliability. This implies that the expression for the reliability of the typical vehicle in the PSP-PPP is sufficient to evaluate that in the OG-PPP,  PLP-PPP, and PLM-PPP by appropriately mapping the system parameters. Also, the vehicular networks behave like PPPs only in the asymptotic regimes of the reliability or data rate. Hence, the street geometry is relevant to understand vehicular network behavior.

An interesting future extension would be to understand how two or more models developed in the Coxian framework can be superimposed to represent an intricate geographic region with intersections and T-junctions, and streets and highways. Also, one may look for tractable models that characterize curved streets and examine whether they are equivalent to the line/stick-based vehicular networks. 

\appendix
\subsection{Proof of Lemma~\ref{hl_approx}}
\label{appendix:hl_approx_proof}
\begin{figure}
\centering
\subfloat[]{\includegraphics[scale=0.58]{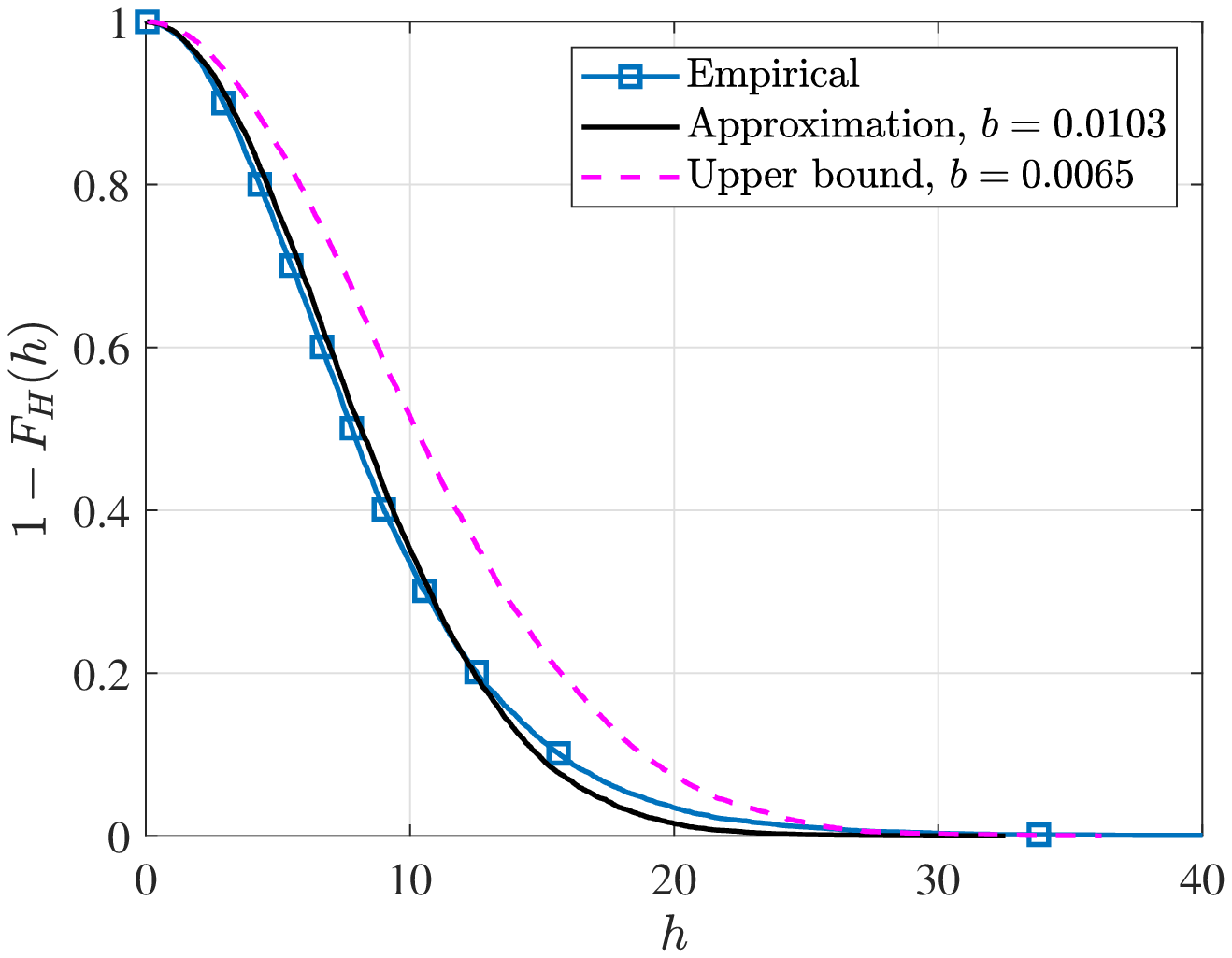} \label{fig:fitted_ccdf}}\hspace{5mm}
\subfloat[]{\includegraphics[scale=0.58]{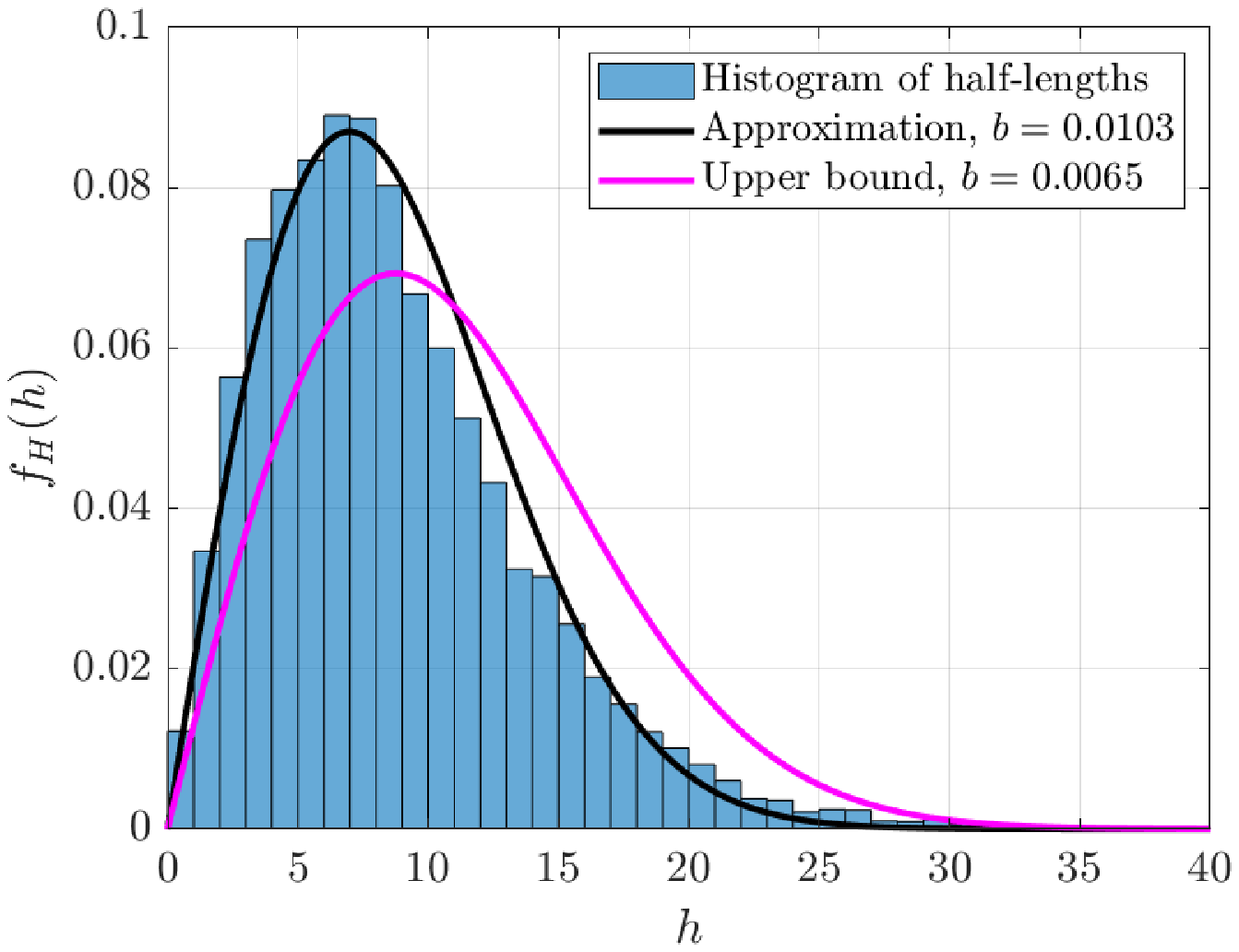}\label{fig:fitted_pdf}}
\caption{Fitting the (a) CCDF and (b) PDF of the half-lengths to that of the Rayleigh distribution. $\mu = 0.01$. }
\label{fig:fitted_ccdf_pdf}
\end{figure}

To prove Lemma~\ref{hl_approx}, we make use of~\cite[Prop. 6.2]{Daley}, which asserts the following:
\textit{There exist $a,b \geq 0$, such that the complementary cumulative distribution function (CCDF) of the half-lengths $H$ in the lilypond model can be bounded as}
\begin{align}
1- F_{H}(h)  \leq a \exp(-b h^{2}), \hspace{3mm} h \geq 0.
\label{eq:hl_theo}
\end{align}
Our goal is to find a tight approximation. By the properties of the cumulative distribution function, $ a \geq 1$. However, $ a > 1$ makes the bound~\eqref{eq:hl_theo} loose for small $h$, which implies that $a = 1 $ is the natural choice. This reduces the bound~\eqref{eq:hl_theo} to the CCDF of the Rayleigh distribution. Fig.~\ref{fig:fitted_ccdf_pdf} fits the CCDF and PDF of the half-lengths to that of the Rayleigh distribution. We observe that the upper bound on the CCDF is loose for small values of $h$. Instead of choosing a value for $b$ that provides an upper bound, we approximate the CCDF of the half-lengths with an appropriate value for $b$ that minimizes the gap between the empirical and theoretical CCDFs. It follows from the Rayleigh approximation that the mean half-length is $\mathbb{E}[H] \approx \sqrt{\pi/4b}$. The value of $b$ is found by equating $\mathbb{E}[H]$ to the empirical mean of the half-lengths.

Next, we study the relation between $b$ and $\mu$. Suppose we scale the PLM by an arbitrary factor $\nu > 0$. Then the street intensity $\mu$ is scaled by $1/\nu^2$ and the half-lengths of the sticks are scaled by $\nu$, inversely proportional to $\sqrt{\mu}$. As the PLM retains its lilypond nature with scaling, the mean of the scaled half-lengths $H'$ is scale-invariant, {\em{i.e.,}} $\sqrt{\pi/4b'} \approx \mathbb{E}[H'] = \mathbb{E}[\nu H] \approx \sqrt{\nu^2\pi/4b}$. This implies that $b' = b/\nu^2$, and thus $b$ scales with $\mu$. 

\setcounter{equation}{26}
\begin{figure*}[b]
\hrule
\begin{align}
{1-F_{R,\mathcal{V}_{o}^{m}}(r)} &= {\mathbb{E} \bigg[\prod \limits_{k = 1}^{ m/2} \mathbb{E}\bigg[ \prod \limits_{z \in {V}_{k} } \mathsf{1}\lbrace{z \notin b(o,r)\rbrace}\bigg]\bigg]}  \stackrel{(a)} = \bigg[\int \limits_{0}^{\infty} \frac{1}{h}\int \limits_{0}^{h} \exp(-\lambda \ell(\gamma, 0, 0) )   \tilde{f}_{H}(h)  \hspace{0.1mm}  \mathrm{d}\gamma  \hspace{0.1mm}  \mathrm{d}h \bigg]^{ m/2 }. \label{eq:lsp_nnf_proof_b} \\
{1-F_{R,\mathcal{V}_{!}}(r)}
& = \exp\bigg(-\frac{\mu}{\pi}\int \limits_{0}^{\infty} \int \limits_{0}^{\pi} \int \limits_{0}^{2\pi} \int \limits_{0}^{r+h}   \exp(-\lambda \ell(\gamma, \phi, \varphi)) \gamma  {f}_{H}(h) \hspace{0.1mm}\mathrm{d}\gamma \hspace{0.1mm}  \mathrm{d}\phi \hspace{0.1mm} \mathrm{d}\varphi \hspace{0.1mm}  \mathrm{d}h \bigg). \label{eq:lsp_nnf_proof_c}
\end{align}
\end{figure*}

\setcounter{equation}{23}
\subsection{Proof of Lemma~\ref{nnf}}
\label{appendix:nnf_decomp_proof}
For the typical vehicle at the origin, the probability that its nearest neighbor is at a distance greater than $r$, $1-F_{R}(r)$, is given by
\begin{align}
1-F_{R}(r) & = \mathbb{E}\bigg[\prod \limits_{z \in \mathcal{V}^{m}} \mathsf{1}\lbrace{z \notin b(o,r)\rbrace}\bigg] \nonumber \\
& \stackrel{(a)} = \mathbb{E}\bigg[\prod \limits_{z \in \mathcal{V}_{o}^{m}} \mathsf{1}\lbrace{z \notin b(o,r)\rbrace}\bigg]\mathbb{E}\bigg[\prod \limits_{z \in \mathcal{V}_{!}} \mathsf{1}\lbrace{z \notin b(o,r)\rbrace}\bigg] \nonumber \\
& = \mathbb{E} \bigg[\prod \limits_{k = 1}^{\lceil m/2 \rceil }  \prod  \limits_{z \in {V}_{k}} \mathsf{1}\lbrace{z \notin b(o,r)\rbrace}\bigg]  \nonumber \\
& \times \mathbb{E} \bigg[\prod \limits_{k > \lceil m/2 \rceil} \prod \limits_{z \in {V}_{k}} \mathsf{1}\lbrace{z \notin b(o,r)\rbrace}\bigg] \nonumber \\
& \stackrel{(b)} = \underbrace{\mathbb{E} \bigg[\prod \limits_{k = 1}^{\lceil m/2 \rceil} \mathbb{E}\bigg[ \prod \limits_{z \in {V}_{k}} \mathsf{1}\lbrace{z \notin b(o,r)\rbrace}\bigg]\bigg]}_{1-F_{R,\mathcal{V}_{o}^{m}}(r)} 
\nonumber \\
& \times  \underbrace{\mathbb{E} \bigg[\prod \limits_{k > \lceil m/2 \rceil} \mathbb{E} \bigg[\prod \limits_{z \in {V}_{k}} \mathsf{1}\lbrace{z \notin b(o,r)\rbrace}\bigg] \bigg]}_{1-F_{R,\mathcal{V}_{!}}(r)}, \label{eq:proof_og_1}
\end{align}
where $(a)$ and $(b)$ exploit the independence of the 1D PPPs.

The contact distance distribution $G_{R}(r)$ is the probability of finding a vehicle within a distance $r$ from an arbitrary location in the plane, say the origin. No streets pass through or contain the origin a.s. This implies that $G_{R}(r)$ is defined with respect to $\mathcal{V}$, which is equivalent in distribution to $\mathcal{V}^{m} \setminus \mathcal{V}_{o}^{m} = \mathcal{V}_{!}$. Hence $G_{R}(r) = F_{R, \mathcal{V}_{!}}(r)$.

\subsection{Proof of Lemma~\ref{nnf_og}}
\label{appendix:nnf_og_proof}
Using the probability generating functional (PGFL) of the PPP, we can express $1-F_{R,\mathcal{V}_{!}}(r)$ in \eqref{eq:proof_og_1} as
\begin{align}
&1-{F}_{R,\mathcal{V}_{!}}(r) \nonumber \\
& = \exp \bigg( \hspace{-2mm}- 2 \mu \int \limits_{0}^{\pi} \int \limits_{0}^{r} (1-\exp(-2 \lambda \sqrt{r^2-u^2})) \mathrm{d}u \hspace{0.3mm}  \mathrm{d} v (\varphi) \bigg), \nonumber \\
& \stackrel{(a)} = \exp \bigg(\hspace{-2mm}- 2 \mu  \int \limits_{0}^{r} (1-e^{-2 \lambda \sqrt{r^2-u^2}}) \mathrm{d}u \bigg) \label{eq:nnf_proof_plp2},
\end{align}
where $(a)$ follows from the independence between $u$ and $\varphi$ and the fact that $\smallint_{\mathbb{R}} \mathrm{d} v (\varphi) = 1$ for the OG-PPP/PLP-PPP. Since the vehicle locations follow a 1D PPP, with respect to the streets that pass through the origin, we have $1-F_{R,\mathcal{V}_{o}^{m}}(r) = (e^{-2 \lambda  r})^{m/2}$.

\subsection{Proof of Theorem~\ref{th_nnf_lsp}}
\label{appendix:nnf_lsp}
The probability that there are no vehicles on a stick $S(y,\varphi,h)$ within a distance $r$ from the origin is $\exp(-\lambda  \vert S(y, \varphi,h) \cap b(o,r) \vert_1)$.  We derive $\vert S(y, \varphi,h) \cap b(o,r) \vert_1$ as follows: The midpoint $y$ is denoted as $(\gamma, \phi)$ in polar coordinates. Using~\eqref{eq:segment}, we can express the points on a stick $ S(y, \phi, h)$ of length $2 h$ as $(\gamma \cos\phi +u \cos\varphi, \gamma \sin\phi+u \sin\varphi)$, $u \in (-h, h)$. Then the squared distance between a point on $S(y, \phi, h)$ and the origin is $\gamma^{2} + u^{2} + 2\gamma u \cos(\phi-\varphi)$. The points of intersection of $S(y,\varphi,h)$ on $b(o,r)$ are at distances $u_{1}, u_{2} = \vert -\gamma \cos(\phi-\varphi) \pm \sqrt{r^2-\gamma^{2}\sin^{2}(\phi-\varphi)} \vert$ from the midpoint of the stick. 
They follow from solving

\begin{equation}
\gamma^{2} + u^{2} + 2\gamma u \cos(\phi-\varphi) = r^{2}
\label{eq:eq_int}
\end{equation}
with respect to $u$. For a stick $S(y, \varphi, h)$ to intersect $b(o,r)$, $y$ must be within $b(o,r+h)$. We consider two cases---$y \in b(o,r)$, and $y \in b(r,r+h)$:
\begin{figure}
\raggedleft
{\includegraphics[scale=0.127]{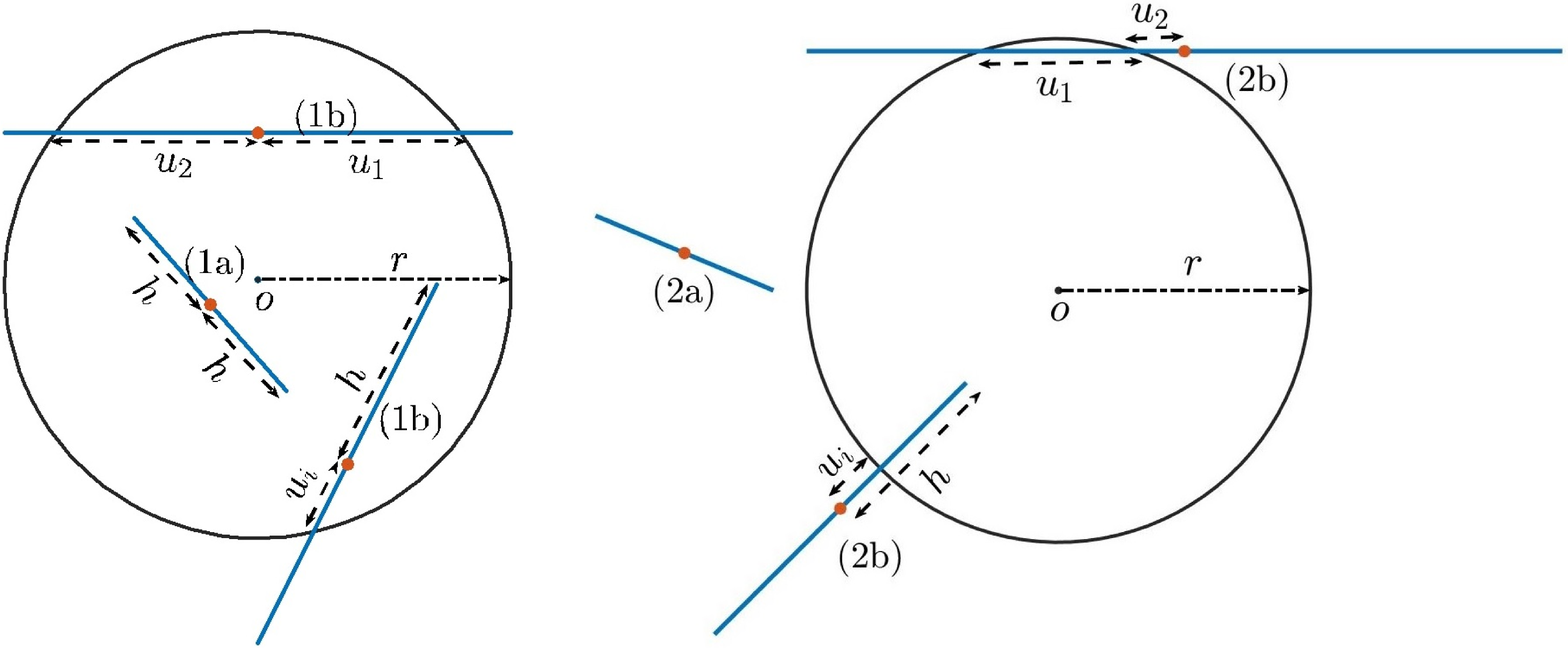} 
\caption{Realizations corresponding to cases (1a), (1b), (2a), and (2b). The midpoints of the streets are highlighted using filled `$\circ$'. The points of intersection of the street on $b(o,r)$ are at distances $u_{i}$, $i \in \lbrace{1,2 \rbrace}$, from the midpoint of the street. $u_{i}$ in the cases (1b) and (2b) refers to $u_{1}$ or $u_{2}$. }
\label{fig:lsp_proof_a}} 
\end{figure}

\noindent {1. \boldsymbol{$y \in b(o,r)$}:} 
Let $ \vert S(y,\varphi,h) \cap b(o,r) \vert \triangleq \ell_{1}$ for $y \in b(o,r) $. When
\begin{itemize}
\item[a.] $S(y,\varphi,h) \cap b(o,r) = S(y, \varphi, h)$: The stick lies within $b(o,r)$. Then $\ell_{1}(\gamma, \phi,\varphi) = 2h$.
\item[b.] $ S(y,\varphi,h) \cap b(o,r) \subset S(y, \varphi, h)$: The stick is not fully contained in $b(o,r)$.  For a stick that passes through $b(o,r)$, $\ell_{1}(\gamma, \phi,\varphi) = u_{1}+u_{2}$. For a stick with only one endpoint in $b(o,r)$, $\ell_{1}(\gamma, \phi,\varphi) = u_{1}+h$ or $ h+u_{2}$.
\end{itemize}
Summarizing the above cases, we write $\ell_{1}(\gamma, \phi,\varphi) =  \min(u_{1}, h) + \min(u_{2}, h)$.

\noindent {2. \boldsymbol{$ y \in b(r, r+h)$}:} 
Let $\vert S(y,\varphi,h) \cap b(o,r) \vert \triangleq \ell_{2}$ for $ y \in b(r, r+h)$. When
\begin{itemize}
\item[a.] $S(y,\varphi,h) \cap b(o,r) = \emptyset$: The stick does not intersect $b(o,r)$. Then $\ell_{2}(\gamma, \phi,\varphi) = 0$.
\item[b.] $S(y,\varphi,h) \cap b(o,r) \subset S(y, \varphi, h)$:  For a stick that passes through $b(o,r)$, $\ell_{2}(\gamma, \phi,\varphi) = \vert u_{1}-u_{2} \vert$. For a stick with only one endpoint in $b(o,r)$, $\ell_{2}(\gamma, \phi,\varphi) = \vert u_{1} -h \vert$ or $ \vert h- u_{2} \vert$.
\end{itemize}
Thus, $\ell_{2}(\gamma, \phi,\varphi) = \vert \min(u_{1}, h) - \min(u_{2} , h)\vert$. Then the probability of not finding a vehicle on $S(y, \varphi,h)$ within $b(o,r)$ is $\exp(-\lambda \ell(\gamma, \phi,\varphi))$, where $\ell(\gamma, \phi,\varphi) = \ell_{1}(\gamma, \phi,\varphi)\mathds{1}_{\gamma \leq r}  + \ell_{2}(\gamma, \phi,\varphi) \mathds{1}_{r < \gamma \leq r+h} $. Fig.~\ref{fig:lsp_proof_a} illustrates the above cases.

For the typical vehicle's street to pass through or end at the origin, its midpoint must belong to $b(o, h)$. The midpoint of the typical vehicle's street of half-length $h$ is at a  distance $\gamma \in [0,h]$ from the origin. Using the above results  and the decomposition of the nearest-neighbor distance distribution in Lemma~\ref{nnf}, we can write ${1-F_{R,\mathcal{V}_{o}^{m}}(r)}$ and ${1-F_{R,\mathcal{V}_{!}}(r)}$ as in \eqref{eq:lsp_nnf_proof_b} and \eqref{eq:lsp_nnf_proof_c}, respectively, 
by applying the PGFL of the PPP. 

Note that $(a)$ in \eqref{eq:lsp_nnf_proof_b} follows from the facts that (i) the rotation of the typical vehicle's streets around the typical vehicle does not change $u_1, u_2$, and hence we can set $\phi = \varphi = 0$, and (ii) $ m/2 $ streets are independent.  Substituting \eqref{eq:lsp_nnf_proof_b} and \eqref{eq:lsp_nnf_proof_c} in \eqref{eq:nnf}, we obtain~\eqref{eq:nnf_lsp}. 

\setcounter{equation}{28}
\subsection{Proof of Proposition~\ref{prop_ps_og_plp}}
\label{appendix:plp}
By~\eqref{eq:line}, we can express the points on $L(x, \varphi)$ as $(x\cos \phi - u\sin\phi, x\sin\phi + u \cos\phi)$, $u \in \mathbb{R}$.  Let $x$ refer to the $k$th-nearest point in $\Phi_1$ from the origin. Then $V_{k}$ denotes the point process of vehicles on $L(x, \phi)$ (Section~\ref{sec:notation}).  The Laplace transform of the interference $\mathcal{L}_{I_{x}}(s)$ from the vehicles on $L(x, \varphi)$ to the typical vehicle at the origin is given by
\begin{align}
\mathcal{L}_{I_{x}}(s)& = \mathbb{E}\bigg[\prod\limits_{z \in V_{k} } \mathbb{E}_{g_z}\left[ \exp\left( -s  g_{z} \Vert z \Vert^{-\alpha} \mathit{B}_z \right) \right] \bigg] \nonumber \\
& = \mathbb{E}\bigg[\prod\limits_{z \in V_{k} : \mathit{B}_{z} =1} \frac{1}{1+s \Vert z \Vert^{-\alpha}} \bigg] \label{eq:ps_basic} \\ 
&  \stackrel{(a)} = \exp\bigg(\hspace{-2mm} - \lambda p \int_{-\infty}^{\infty}  \frac{1}{1+\left(\frac{x^{2}+ u^2}{s^\delta} \right)^{1/\delta}} {\mathrm{d}u} \bigg)  \nonumber\\ 
& \stackrel{(b)}  = \exp \bigg(\hspace{-2mm}-  \lambda p s^{\delta/2} \int_{v_x}^{\infty} \frac{1}{\left( 1 + v^{{1/\delta}}\right) \sqrt{v - v_x}} \mathrm{d}v\bigg), \label{eq:ps_distant_axes} 
\end{align}
where $\delta = 2/\alpha$, $(a)$ substitutes $\Vert z \Vert _{2}= \Vert (x\cos \phi - u\sin\phi, x\sin\phi + u \cos\phi) \Vert_{2}$ and the PGFL of the PPP. $\lambda p$ is the intensity of the active transmitters for which $\mathit{B}_{z} =1$, $z \in V_{k}$. $(b)$ is due to the change of variable $v = \frac{x^{2} + u^{2}}{s^\delta}$, and $v_x = \frac{x^2}{s^{\delta}}$. 

We learned from \eqref{eq:sir} that the success probability is the product of the Laplace transforms of the interference $I_{o}^{m}$ and $I_{!}$, which we derive below.

\setcounter{equation}{40}
\begin{figure*}[b]
\hrule
\begin{align}
& \mathcal{L}_{I_{!}}^{\mathrm{PSP-PPP}}(\theta D^{\alpha}) \sim 1- \frac{\theta \mu \lambda p}{\pi} \int\limits_{0}^{\infty} \int\limits_{0}^{\pi} \int\limits_{0}^{2 \pi} \int\limits_{0}^{\infty}  \int\limits_{-h}^{h} \bigg(\frac{ D^{2}}{{\gamma^2+u^2+2 \gamma u \cos(\phi-\varphi)}}\bigg)^{1/\delta}  \gamma {f}_{H}(h) \mathrm{d}u \hspace{0.1mm} \mathrm{d}\gamma \hspace{0.1mm} \mathrm{d}\phi \hspace{0.1mm} \mathrm{d}\varphi \hspace{0.1mm} \mathrm{d}h  = 1 - \Theta(\theta). \label{eq:LIr} \\
& \lim_{\theta \to 0} \hspace{1mm} 1 - p_{m}^{\mathrm{PSP-PPP}}  = 1-\lim_{\theta \to 0} \hspace{1mm} \mathbb{E}\bigg[\exp \Bigg(- \lambda p  s^{\delta/2} \int\limits_{(-W-H) s^{-\delta/2}}^{(-W+H) s^{-\delta/2}} \frac{1}{ 1 + v^{{2/\delta}}} \mathrm{d}v\bigg) \Bigg]^{m/2} \label{eq:term2} \\
& \hspace{28mm} = 1 -  \mathbb{E}\Bigg[1- \lambda p  s^{\delta/2} \int\limits_{(-W-H) s^{-\delta/2}}^{(-W+H) s^{-\delta/2}} \frac{1}{ 1 + v^{{2/\delta}}} \mathrm{d}v \Bigg]^{m/2} \label{eq:term0} \\
& \hspace{28mm}  = 1 - \mathbb{E}\Bigg[1- \frac{\lambda p  s^{\delta/2}}{2}  \int\limits_{(-W-H)^{2} s^{-\delta/2}}^{(-W+H)^{2} s^{-\delta/2}} \frac{1}{ (1 + u^{{1/\delta}}) \sqrt{u}} \hspace{0.3mm} \mathrm{d}u \Bigg]^{m/2} \label{eq:term1} \sim K_{\alpha} \lambda p  s^{\delta m/4}.
\end{align}
\end{figure*}

\setcounter{equation}{30}
\subsubsection{Laplace Transform of $I_{o}^{m}$}
The Laplace transform of the interference from $V_{0}$ to the typical general vehicle (order 2) can be obtained by setting $x=0$ in~\eqref{eq:ps_distant_axes}, {\em{i.e.,}} 
\begin{align}
\mathcal{L}_{I_{o}^{2}}(s)& = \exp(- 2\lambda p s^{\delta/2}\Gamma(1+\delta/2)\Gamma(1-\delta/2)). \label{eq:ps_same_axes}
\end{align}
For the typical intersection vehicle (order 4), as two streets pass through the origin,  $\mathcal{V}_{o}^{4} = V_0 \cup V_1$ (Sec.~\ref{sec:notation}). Then $ {\mathcal{L}}_{I_{o}^{4}}(s)$ can be written as in~\eqref{eq:ps_basic} as follows:

\begin{align}
 {\mathcal{L}}_{I_{o}^{4}}(s) &= \mathbb{E}\bigg[\prod\limits_{z \in V_0 \cup V_{1}: \mathit{B}_{z} =1} \frac{1}{1+s \Vert z \Vert^{-\alpha}}   \bigg] \nonumber \\
& \stackrel{(c)} = \bigg( \mathbb{E}\Bigg[\prod\limits_{z \in V_0: \mathit{B}_{z} =1} \frac{1}{1+s \Vert z \Vert^{-\alpha} } \bigg] \bigg)^{2},
\label{eq:ps_ti}
\end{align}
where $(c)$ results from $V_{0}$ being identically distributed as $V_{1}$. It follows that $\mathcal{L}_{I_{o}^{4}}(s) = \mathcal{L}^{2}_{I_{o}^{2}}(s)$.
\subsubsection{Laplace Transform of $I_{!}$}
 The aggregate Laplace transform of the interference from the vehicles on all the streets that do not pass through the typical vehicle is given by
\begin{align}
\mathcal{L}_{I_{!}}(s) &= \mathbb{E}\bigg[\prod\limits_{z \in \mathcal{V}_{!}} \mathbb{E}_{g_z}\left[\exp\left( -s  g_{z} \Vert z \Vert^{-\alpha} \mathit{B}_{z} \right) \right] \bigg] \nonumber \\
 & \stackrel{(e)}=\mathbb{E}\bigg[\prod \limits_{k > m/2} \mathbb{E}\bigg[\prod\limits_{z \in V_{k}: \mathit{B}_{z} =1} \frac{1}{1+s \Vert z \Vert^{-\alpha} } \bigg]\bigg] \label{eq:proof_basic2} \\
 & \stackrel{(f)} = \exp\bigg(\hspace{-2mm}-  \mu \int_{0}^{\pi} \int_{-\infty}^{\infty} (1-\mathcal{L}_{I_{x}}(s)) \hspace{0.5mm} \mathrm{d}x \hspace{0.5mm} \mathrm{d} \nu(\varphi) \bigg) \nonumber \\
 & \stackrel{(g)} = \exp\bigg(\hspace{-2mm}-  \mu  \int_{-\infty}^{\infty} (1-\mathcal{L}_{I_{x}}(s)) \hspace{0.5mm}  \mathrm{d}x \bigg),\label{eq:ps_diff}
\end{align}
where $(e)$ follows from $\mathcal{V}_{!} = \cup_{k > m/2} V_{k}$ and the fact that the $V_{k}$'s are independent 1D PPPs, $(f)$ uses~\eqref{eq:ps_distant_axes} and the PGFL of the PPP with respect to $x$ and $\varphi$, and $(g)$ results from $\mathcal{L}_{I_{x}}(s)$ being independent of $\varphi$, and $\int_{\mathbb{R}} \mathrm{d}\nu(\varphi) = 1$. Substituting \eqref{eq:ps_same_axes}, \eqref{eq:ps_ti}, and \eqref{eq:ps_diff} in \eqref{eq:sir}, we obtain~\eqref{eq:ps_og_ppp}. 
 
\subsection{Proof of Proposition~\ref{prop_lio_psp}}
\label{appendix:lio_psp}
Using~\eqref{eq:segment}, we can denote the points on a stick $S(y, \phi, h) $ of length $2 h$ as $(\gamma \cos\phi +u \cos\varphi, \gamma \sin\phi$ $ + u \sin\varphi)$, $u \in (-h, h)$. The midpoint of the street that passes through the typical vehicle is at a signed distance $W \sim \mathcal{U}(-h,h)$ from the origin. Then the endpoints of the street are at signed distances $-W-h$, and $-W+h$ to the origin. As in~\eqref{eq:ps_basic}, the Laplace transform of the interference $I_{o}^{2}$ for the typical general vehicle can be written as
\begin{align}
\mathcal{L}_{I_{o}^{2}}(s)& = \mathbb{E}\bigg[\prod\limits_{z \in \mathcal{V}_{o}: \mathit{B}_{z} =1} \frac{1}{1+s \Vert z \Vert^{-\alpha} } \bigg] \nonumber  \\
& \stackrel{(a)} = \mathbb{E}\bigg[\exp\bigg(\hspace{-2mm} -\lambda p \int \limits_{-W-H}^{-W+H} \frac{1}{1+\left(\frac{u}{s^{\delta/2}} \right)^{2/\delta}} {\mathrm{d}u} \bigg)\bigg]   \label{eq:exp_psp_proof} \\
& \stackrel{(b)} = \mathbb{E}\bigg[\exp \bigg(\hspace{-2mm}- \lambda p  s^{\delta/2} \int \limits_{\frac{(-W-H)}{ s^{\delta/2}}}^{\frac{(-W+H)}{ s^{\delta/2}}} \frac{1}{ 1 + v^{{2/\delta}}} \mathrm{d}v\bigg)\bigg], \label{eq:psp_proof_hr} 
\end{align}
where $(a)$ substitutes $\Vert z \Vert_2 = \Vert (u\cos\varphi, u\sin\varphi) \Vert_{2}$ and applies the PGFL of the PPP, and $(b)$ results from the change of variable $ v = \frac{u}{s^{\delta/2}}$. $(c)$ evaluates the expectation with respect to $H$ using $\tilde{f}_{H}(h)$ rather than ${f}_{H}(h)$ based on Lemma~\ref{lemma_ins_paradox}. For the typical intersection vehicle, as two streets pass through the origin and the point processes on them are independent, the corresponding Laplace transform of the interference is 
the square of that given in \eqref{eq:psp_proof_hr} similar to \eqref{eq:ps_ti}.

\subsection{Proof of Proposition~\ref{prop_lir_psp}}
\label{appendix:lir_psp}
This proof uses the same notation as in Appendix~\ref{appendix:lio_psp}. Let $\mathcal{A} \triangleq \mathbb{R}^{+} \times [0, 2\pi) \times [0, \pi) \times \mathbb{R}^{+}$ and $a = (\gamma,\phi,\varphi,h) \in \mathcal{A}$.  From Section~\ref{sec:notation}, we have $V_{k}$, $k > m/2$, denoting the point process of vehicles on the $k$th street $S(\gamma,\phi,\varphi,h)$ that does not pass through the origin. The Laplace transform of the interference $\mathcal{L}_{I_{a}}(s)$ from the vehicles on $S(\gamma,\phi,\varphi,h)$ is 
\begin{align}
\mathcal{L}_{I_{a}}(s) &= \mathbb{E}\bigg[\prod\limits_{z \in V_{k}: \mathit{B}_{z} =1} \frac{1}{1+s \Vert z \Vert^{-\alpha} } \bigg] \nonumber \\
& \stackrel{(a)} = \exp\bigg(\hspace{-2mm}-\lambda p \int \limits_{-h}^{h} \frac{1}{1+\frac{{\mathfrak{F}(\gamma, u, \phi, \varphi)}^{1/\delta}}{s}} \mathrm{d}u \bigg), \label{eq:ps_arb_psp}
\end{align}
where $(a)$ applies the PGFL of the PPP, $\Vert z \Vert_{2} = \Vert(\gamma \cos\phi +u \cos\varphi, \gamma \sin\phi+u \sin\varphi)\Vert_{2}$, and $\mathfrak{F}(\gamma, u, \phi, \varphi) = \gamma^2+u^2+2\gamma u \cos(\phi-\varphi)$.

Using \eqref{eq:proof_basic2}, \eqref{eq:ps_arb_psp}, and applying PGFL with respect to midpoints, orientations and half-lengths, we write the Laplace transform of the interference from $\mathcal{V}_{!}$ as
\begin{align}
\mathcal{L}_{I_{!}}(s) & =\mathbb{E}\bigg[\prod \limits_{k > m/2} \mathbb{E}\bigg[\prod\limits_{z \in V_{k}: \mathit{B}_{z} =1} \frac{1}{1+s \Vert z \Vert^{-\alpha} } \bigg]\bigg] \nonumber \\
&  = \exp\bigg(\hspace{-2mm}-\frac{\mu}{\pi} \int_{\mathcal{A}} (1-\mathcal{L}_{I_{a}}(s))  \gamma {f}_{H}(h) \mathrm{d}\mathcal{A} \bigg), \label{eq:ps_diff_proof}
\end{align}
where $\mathrm{d}\mathcal{A} = \mathrm{d}\gamma \hspace{0.3mm} \mathrm{d}\phi \hspace{0.3mm} \mathrm{d}\varphi \hspace{0.3mm} \mathrm{d}h$. 

\subsection{Proof of Proposition~\ref{lem_lp_ppp}}
\label{appendix:lp}
\subsubsection{Success probability of the typical general vehicle in the PLM-PPP}
We learned from Conjecture 1 that the probability of finding the $n$th-nearest neighbor closer is higher in the PSP-PPP than in the PLM-PPP. Consequently, the success probability of the typical general vehicle in the PLM-PPP is lower bounded by that in the PSP-PPP. As this inference is based on a conjecture, we present the result for the success probability of the typical general vehicle in the PLM-PPP as an approximation rather than a bound. By~\eqref{eq:sir_joint} and \eqref{eq:sir}, we have
\begin{align}
p_{2}^\mathrm{PLM-PPP} & =  \mathcal{L}_{I_{o}^{2} + I_{!}}^{\mathrm{PLM-PPP}}(s) \nonumber \\
&\approx \mathcal{L}_{I_{o}^{2}}^{\mathrm{PSP-PPP}}(s)\mathcal{L}_{I_{!}}^{\mathrm{PSP-PPP}}(s) \mid_{ f_{H}(h) =  \hat{f}_{H}(h)}.
\label{eq:sir_lp}
\end{align}

\subsubsection{Success probability of the typical T-junction vehicle in the PLM-PPP}
We have $\mathcal{V}_{o}^{3} = V_0 \cup V_1$, where $V_{0}$ denote the vehicles on the street that passes through the origin, and $V_{1}$ denote the vehicles on the street with one endpoint at the T-junction. The success probability of the typical T-junction vehicle (order 3), $p_{3}^\mathrm{PLM-PPP}$, is given by
\begin{align}
p_{3} &= \mathbb{E}\Bigg[\prod\limits_{\substack{z \in V_0 \cup V_1: \\ \mathit{B}_{z} =1}} \frac{1}{1+s \Vert z \Vert^{-\alpha} }  \prod\limits_{\substack{z \in \mathcal{V}_{!}: \\ \mathit{B}_{z} =1}}  \frac{1}{1+s \Vert z \Vert^{-\alpha} }  \Bigg] \nonumber \\
& \stackrel{(a)} \approx   \mathbb{E}\Bigg[\prod\limits_{\substack{z \in V_{0} \cup \mathcal{V}_{!}: \\ \mathit{B}_{z} =1}} \frac{1}{1+s \Vert z \Vert^{-\alpha} }  \Bigg] \mathbb{E}\Bigg[\prod\limits_{\substack{z \in V_1: \\ \mathit{B}_{z} =1}} \frac{1}{1+s \Vert z \Vert^{-\alpha} } \Bigg]\nonumber \\
& \stackrel{(b)} \approx \mathcal{L}_{I_{o}^{2}}^{\mathrm{PSP-PPP}}(s)\mathcal{L}_{I_{!}}^{\mathrm{PSP-PPP}}(s) \hspace{0.3mm}\mathbb{E}\Bigg[\prod\limits_{\substack{z \in V_1: \\ \mathit{B}_{z} =1}} \frac{1}{1+s \Vert z \Vert^{-\alpha} } \Bigg] \nonumber \\
& \stackrel{(c)} =  \mathcal{L}_{I_{o}^{2}}^{\mathrm{PSP-PPP}}(s)\mathcal{L}_{I_{!}}^{\mathrm{PSP-PPP}}(s)  \nonumber \\
& \times  \int \limits_{0}^{\infty} \exp \bigg(\hspace{-2mm}- \lambda p  s^{\delta/2} \int \limits_{0}^{ 2h s^{-\delta/2}} \frac{1}{ 1 + v^{{2/\delta}}} \mathrm{d}v\bigg) \hat{f}_{H}(h) \mathrm{d}h , \label{eq:ps_same_tjn}
\end{align}
where $(a)$ assumes independence between $V_{1}$ and $V_{0} \cup \mathcal{V}_{!}$,  and $(b)$ follows from~\eqref{eq:sir_lp}. The integral expression in $(c)$ can be derived similarly to~\eqref{eq:psp_proof_hr} using the detail that for the street whose one endpoint is a T-junction at the origin, its other endpoint is at a distance $2H$ from the origin. $\hat{f}_{H}(h)$ is the approximated density of the half-length of the street that ends at a T-junction.

\subsection{Proof of Lemma~\ref{th_lsp_zero}}
\label{appendix:th_lsp_zero}
We can approximate $\mathcal{L}_{I_{!}}(\theta D^{\alpha})$ given by~\eqref{eq:ps_diff_proof} using Taylor's series as $\theta \to 0$ as in~\eqref{eq:LIr}. 
Using~\eqref{eq:exp_psp_proof} and \eqref{eq:LIr}, the outage probability as $\theta \to 0$, can be written as in~\eqref{eq:term2}, where $s = \theta D^{\alpha}$. By interchanging the limit and expectation (using the monotone convergence theorem), and applying Taylor's theorem inside the expectation in~\eqref{eq:term2}, we obtain~\eqref{eq:term0}. It transforms to~\eqref{eq:term1} by the change of variable $u = v^{2}$, whose integral term evaluates to a constant $K_{\alpha}$ as $\theta \to 0$. Note that $K_{\alpha}$ depends on $\alpha$.

\setcounter{equation}{44}
\subsection{Proof of Lemma~\ref{th_lsp_infty}}
\label{appendix:th_lsp_infty}
Consider a model $\mathcal{V}'$ formed by mapping all the points on each stick to its midpoint. 
The expected number of active transmitters on each stick is $2 \lambda p \mathbb{E}[H]$. Since the midpoints follow a 2D PPP of intensity $\mu$, $\mathcal{V}'$ forms a non-simple 2D PPP 
with density $2 \mu \lambda p \mathbb{E}[H]$. It can be seen as a marked point process whose ground process is a 2D PPP of intensity $\mu$ and the marks $2 \lambda p \mathbb{E}[H]$ refer to the multiplicity of the points. 
Let $\mathbb{M}$ define the mapping from $\mathcal{V}$ to $\mathcal{V}'$. Using~\eqref{eq:sir_def} and \eqref{eq:sir_expr}, the success probability of the typical vehicle in the PSP-PPP can be expressed as
\begin{equation}
p_{m}^{\mathrm{PSP-PPP}} = \mathbb{P}\bigg( g > D^{\alpha} \sum \limits_{z \in \mathcal{V} } g_{z} \Vert \theta^{-1/\alpha} z \Vert^{-\alpha} B_{z} \bigg). \nonumber
\end{equation}
Similarly, the success probability of the typical vehicle in $\mathcal{V}'$ is written as
\begin{equation}
{p_{m}^{\mathcal{V}'}} = \mathbb{P}\bigg( g > D^{\alpha} \sum \limits_{z \in \mathcal{V} } g_{z} \Vert \theta^{-1/\alpha} \mathbb{M}(z) \Vert^{-\alpha} B_{z}\bigg). \nonumber \\
\end{equation}
For sticks of half-length $h$, $\vert \Vert z \Vert - \Vert \mathbb{M}(z) \Vert \vert \leq \Vert z-\mathbb{M}(z) \Vert \leq h < \infty$. Then
\begin{equation}
\vert \Vert \theta^{-1/\alpha} z \Vert - \Vert \theta^{-1/\alpha} \mathbb{M}(z) \Vert \vert \to 0 \hspace{3mm} \text{as} \hspace{3mm}  \theta \to \infty. 
\end{equation}
Then $p_{m}^{\mathrm{PSP-PPP}} \to p_{m}^{\mathcal{V}'}$ as $\theta \to \infty$. The order $m$ of the typical vehicle is irrelevant as the mapping $\mathbb{M}$ does not distinguish an intersection from a non-intersection. 
The success probability of the typical vehicle $p_{m}^{\mathcal{V}'}$ can be obtained by substituting $\lambda_{2} = 2 \mu \lambda p \mathbb{E}[H]$ in \eqref{eq:ps_nd_ppp}.
\bibliographystyle{IEEEtran}
\bibliography{references}

\begin{IEEEbiography}[{\includegraphics[width=1in,height=1.25in,clip,keepaspectratio]{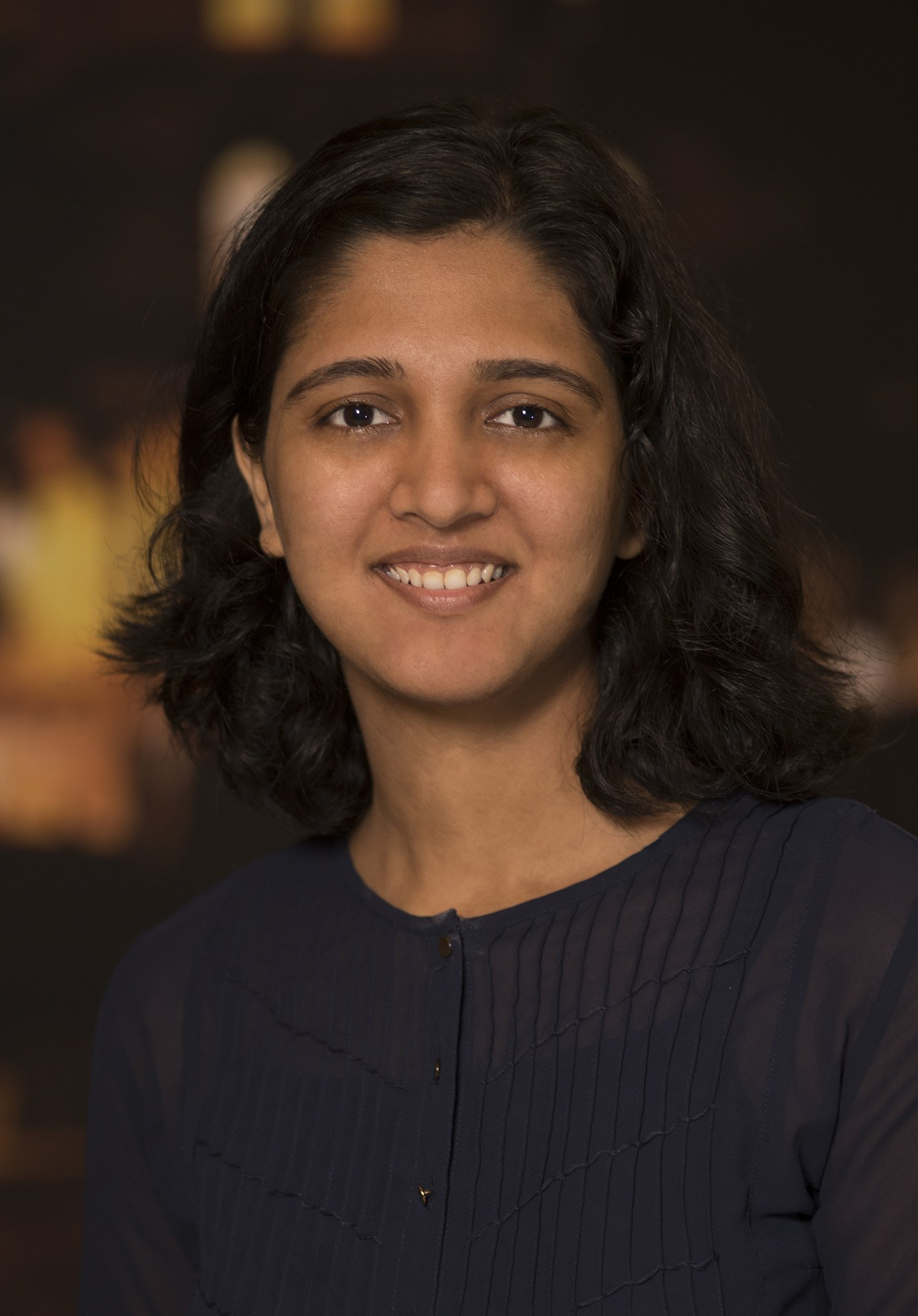}}]{Jeya Pradha Jeyaraj}
(S'17) received the M.Tech degree in electrical engineering from the Indian Institute of Technology Kanpur, India, in 2013. She is currently pursuing the Ph.D. degree in electrical engineering at the University of Notre Dame, Indiana, USA, since 2016. She is also working towards the M.S. degree in applied and computational mathematics and statistics. Before joining her Ph.D., she worked as a Software Development Engineer in the 3G cellular technologies group at Cisco Systems, India. She held internship positions in the research and development divisions at Toyota Motor North America and Qualcomm in the USA. Her research interests include vehicular networks, stochastic geometry, spectrum sharing, and data mining. 
\end{IEEEbiography}

\begin{IEEEbiography}
  [{\includegraphics[width=1in,height=1.25in,clip,keepaspectratio]{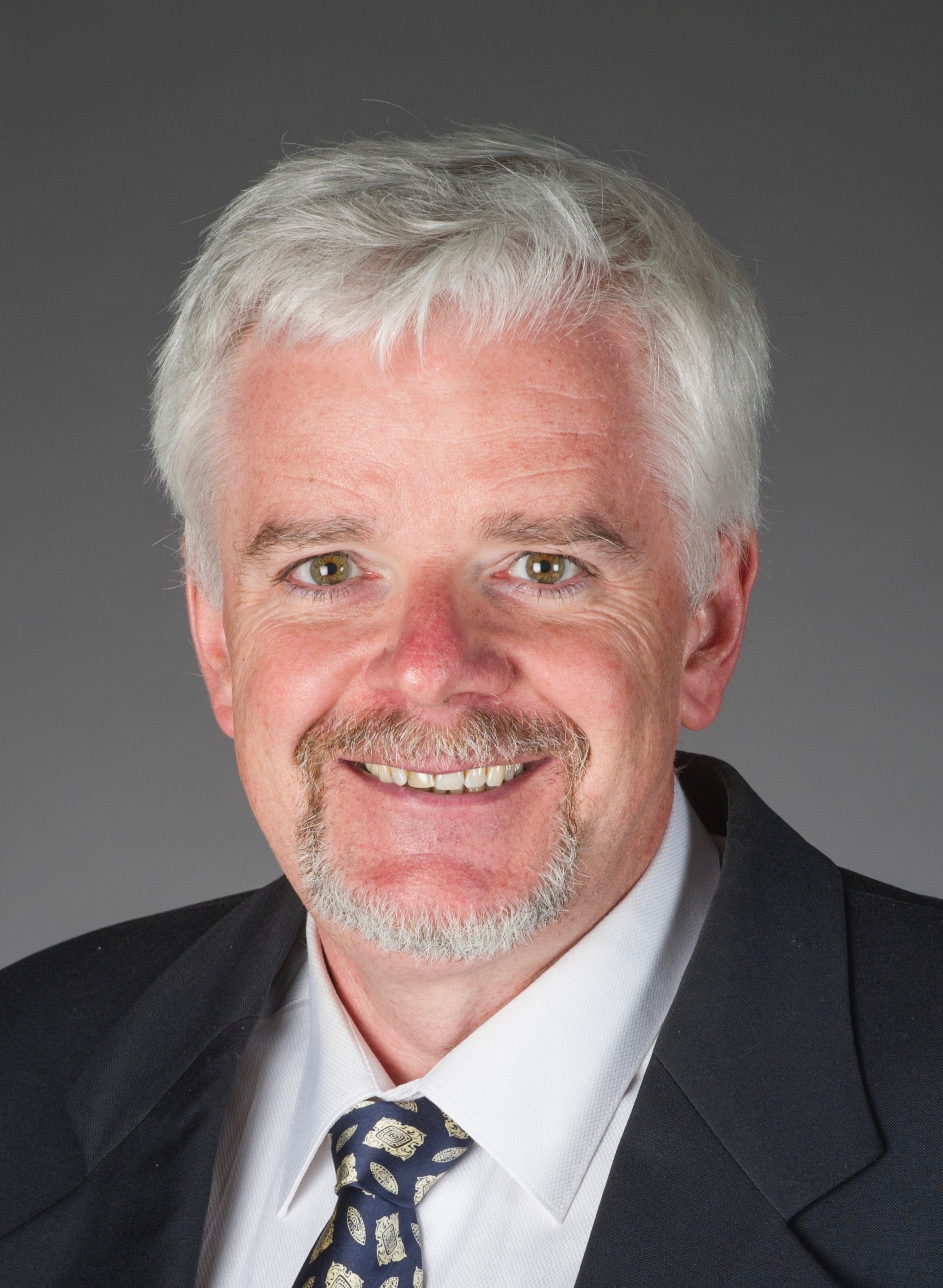}}]{Martin Haenggi}
 (S`95-M`99-SM`04-F`14) received the Dipl.-Ing. (M.Sc.) and Dr.sc.techn. (Ph.D.) degrees in electrical engineering from the Swiss Federal Institute of Technology in Zurich (ETH) in 1995 and 1999, respectively. Currently he is the Freimann Professor of Electrical Engineering and a Concurrent Professor of Applied and Computational Mathematics and Statistics at the University of Notre Dame, Indiana, USA. In 2007-2008, he was a visiting professor at the University of California at San Diego, and in 2014-2015 he was an Invited Professor at EPFL, Switzerland.
He is a co-author of the monographs "Interference in Large Wireless Networks" (NOW Publishers, 2009) and `Stochastic Geometry Analysis of Cellular Networks' (Cambridge University Press, 2018) and the author of the textbook "Stochastic Geometry for Wireless Networks" (Cambridge, 2012), and he published 16 single-author journal articles. His scientific interests lie in networking and wireless communications, with an emphasis on cellular, amorphous, ad hoc (including D2D and M2M), cognitive, and vehicular networks.
He served as an Associate Editor of the Elsevier Journal of Ad Hoc Networks, the IEEE Transactions on Mobile Computing (TMC), the ACM Transactions on Sensor Networks, as a Guest Editor for the IEEE Journal on Selected Areas in Communications, the IEEE Transactions on Vehicular Technology, and the EURASIP Journal on Wireless Communications and Networking, as a Steering Committee member of the TMC, and as the Chair of the Executive Editorial Committee of the IEEE Transactions on Wireless Communications (TWC). From 2017 to 2018, he was the Editor-in-Chief of the TWC. Currently he is an editor for MDPI Information. For both his M.Sc. and Ph.D. theses, he was awarded the ETH medal. He also received a CAREER award from the U.S. National Science Foundation in 2005 and three awards from the IEEE Communications Society, the 2010 Best Tutorial Paper award, the 2017 Stephen O. Rice Prize paper award, and the 2017 Best Survey paper award, and he is a Clarivate Analytics Highly Cited Researcher.
\end{IEEEbiography}

\end{document}